# Pulsed laser deposition of single phase n- and p-type Cu$_2$O thin films with low resistivity

*Syed Farid Uddin Farhad[1, 2, 3, 4*], David Cherns[1], James Smith[2], Neil Fox[1, 2], and David Fermín[3]*

[1]H.H. Wills Physics Laboratory, School of Physics, University of Bristol, BS8 1TL, UK

[2]Diamond Laboratory, School of Chemistry, University of Bristol, BS8 1TS, UK

[3]Electrochemistry Laboratory, School of Chemistry, University of Bristol, BS8 1TS, UK

[4]Industrial Physics Division, BCSIR Laboratories, Dhaka, Bangladesh Council of Scientific and Industrial Research (BCSIR), Dhaka 1205, Bangladesh

[*]Corresponding Author: sf1878@my.bristol.ac.uk , s.f.u.farhad@bcsir.gov.bd

**Abstract**

Low resistivity ($\rho \sim$ 3-24 m$\Omega$.cm) with tunable n- and p-type single phase Cu$_2$O thin films have been grown by pulsed laser deposition at 25-200 $^0$C by varying the background oxygen partial pressure (O$_{2pp}$). Capacitance data obtained by electrochemical impedance spectroscopy was used to determine the conductivity (n- or p-type), carrier density, and flat band potentials for samples grown on indium tin oxide (ITO) at 25 $^0$C. The Hall mobility ($\mu_H$) of the n- and p-type Cu$_2$O was estimated to be $\sim$ 0.85 cm$^2$.V$^{-1}$s$^{-1}$ and $\sim$ 4.78 cm$^2$.V$^{-1}$s$^{-1}$ respectively for samples grown on quartz substrate at 25 $^0$C. An elevated substrate temperature $\sim$ 200 $^0$C with O$_{2pp}$ = 2 - 3 mTorr yielded p-type Cu$_2$O films with six orders of magnitude higher resistivities in the range, $\rho \sim$ 9 - 49 k$\Omega$.cm and mobilities in the range, $\mu_H \sim$ 13.5 - 22.2 cm$^2$.V$^{-1}$s$^{-1}$. UV-Vis-NIR diffuse reflectance spectroscopy showed optical bandgaps of Cu$_2$O films in the range of 1.76 to 2.15 eV depending on O$_{2pp}$. Thin films grown at oxygen rich conditions O$_{2pp} \geq$ 7 mTorr yielded mixed phase copper oxide irrespective of the substrate temperatures and upon air annealing at 550 $^0$C for 1 hour completely converted to CuO phase with n-type semiconducting properties ($\rho \sim$12 $\Omega$.cm, $\mu_H \sim$1.50 cm$^2$V$^{-1}$s$^{-1}$). The as-grown p- and n-type Cu$_2$O showed rectification and a photovoltaic response in solid junctions with n-ZnO and p-Si electrodes respectively. Our findings may create new opportunities for devising Cu$_2$O based junctions requiring low process temperatures.

**Keywords**: Pulsed Laser Deposition; Cuprous oxide (Cu$_2$O) thin film; p- and n-type conductivity; Hall coefficient measurement; Fermi level; Mott-Schottky analyses.

1. **Introduction**

The semiconductor cuprous oxide ($Cu_2O$), has shown much promise in photocatalytic water splitting [1], resistive switching devices [2], thin film transistors (TFT) [3], gas sensors [4], as an anode material in Li-ion based batteries [5] and in photovoltaic (PV) applications [6]. It is non-toxic, Earth-abundant, and could be prepared by physical and chemical methods [6-8]. In addition, single phase $Cu_2O$ is desirable as an absorber material for solar cells because of its reported direct bandgap ($E_g$~2.17 eV) and high absorption coefficient (above $10^5$ cm$^{-1}$) in the visible region of solar radiation and, potentially, the ability to dope both n- and p-type via manipulating processing conditions [9-15]. The natural p-type conductivity of $Cu_2O$ is believed to stem from copper vacancies ($V_{Cu}$) in the crystal lattice [16, 17], and its suitable band alignment with other wide bandgap n-type metal oxide semiconductors, such as ZnO [18, 19] and $TiO_2$ [20, 21], makes it attractive for realizing reasonably efficient heterojunction solar cells. In contrast, the origin of intrinsic n-type conductivity is a matter of debate [9, 22, 23], yet it has gained substantial interest because of the possibility of the formation of homojunction solar cells [7, 9, 12, 14, 24] to suppress the deleterious interfacial defects states often formed in the case of most heterojunction solar cells ([25] and refs. therein). The oxygen vacancy ($V_o$) [11] and copper interstitial ($Cu_i$) defects in $Cu_2O$ lattice have been proposed for the electron-donating source in explaining the experimental results ([22] and refs. therein). However, the formation energy of $V_o$ is relatively low compared to that of $Cu_i$ irrespective of the copper-rich (poor)/oxygen-poor (rich) growth conditions [22]. In general, the donor- and acceptor- levels should be shallow for effective n- and p- type semiconductors. Carrier concentrations over of ~$10^{16}$ cm$^{-3}$ and mobilities at least ~5 cm$^2$/V.s are also necessary for efficient $Cu_2O$ based optoelectronic devices [11, 26, 27]. In the case of intrinsic n-type $Cu_2O$, most of the experimental works reported deep donor levels ([22] and refs. therein), therefore, unlikely to overcome the native p-type conductivity stemming from cation deficiency ($V_{Cu}$) [6] due to self-compensation [28]. However, a recent report by Nandy et al. demonstrated that controlling the oxygen vacancies ($V_o$) into the into the $Cu_2O$ crystal (i.e., $Cu_2O_{1-\delta}$) may induce an impurity state close to the conduction band thereby enhancing the electron-donating ability due to the unshared d-electrons of Cu atoms (nearest to the vacancy site) [29]. Their theoretical approach further suggested that the formation energy for obtaining 2.08% ($\delta$=0.208) of $V_o$ in $Cu_2O$ could be as low as 0.32 eV under oxygen deficient condition, suggesting that the stable $Cu_2O_{1-\delta}$ entity maybe experimentally viable [29].

Several physical [6, 25, 30-32] and wet-chemical [3, 5, 15, 33-36] based deposition methods have been employed to produce phase pure $Cu_2O$. Among them, pulsed laser deposition (PLD) offers good control on a wide range of processing parameters such as laser wavelength, pulse repetition rate, laser energy per pulse (LP), background gas pressure ($O_{2pp}$), substrate temperature ($T_{sub}$) etc. and has been demonstrated to give single phase cuprous oxides with good structural, morphological and electrical properties [11, 25, 27, 37, 38]. Previous work [25, 37] demonstrated the similar microstructures of $Cu_xO_y$ (a defect variant of the basic-$Cu_2O$) thin films on various substrates using different PLD processing conditions, these identified favorable process conditions within which we now focus. These investigations, on the effect of $T_{sub}$ ($\leq 400\ ^0C$) and $O_{2pp}$ ($\leq 10$ mTorr), achieve single phase $Cu_2O$ with varying optical and electrical properties by exploiting the non-equilibrium deposition environment of PLD. The ablation species react with ambient background gas chemically and physically while it proceeds towards the substrate from the target and affects the resulting film properties. Oxygen content in copper oxide films can be adjusted through controlling the oxygen partial pressure ($O_{2pp}$) during growth. In the literature most of the investigations related to electrical properties of Cu-O films deal with films deposited under oxygen rich conditions near the $Cu_2O$/CuO boundary. There is less information for films deposited at the Cu/$Cu_2O$ boundary, where an n-type $Cu_2O$ may exist [39] due to the predominant presence of oxygen vacancy ($V_o$) as opposed to the hole creating copper vacancy ($V_{cu}$). Meyer et al. [6], roughly estimated the binding energy of donor and acceptors for $Cu_2O$ to be 266 meV and 156 meV using effective mass theory (EMT) and discussed that carrier densities well above $\sim 10^{18}$ cm$^{-3}$ are necessary to overcome natural p-type conductivity due to the cation vacancy ($V_{cu}$). Wang et al. [40] argued that n-type conductivity of $Cu_2O$ is possible only if the $V_o$ would exist at very high concentrations synthesized in oxygen poor conditions. In our previous report we demonstrated the n-type conductivity of $Cu_2O$ based on electrochemical Mott-Schottky analyses for the samples grown by PLD at room temperature with oxygen poor condition ($O_{2pp} << 1$ mTorr) [41]. Recently, Xu et al. [11], also reported PLD grown phase pure $Cu_2O$ films produced at 600 $^0$C under oxygen poor conditions ($O_{2pp}$ = 0.09 Pa $\approx$ 0.68 mTorr). The films exhibited p-type conductivity and upon post $N_2$ plasma treatment of the as-deposited samples they observed a phase transition from pure $Cu_2O$ to a mixture of $Cu_2O$ and Cu with an accompanied change from p to n-type conduction. Therefore, in this work, the effects of $O_{2pp}$ (both oxygen-rich and oxygen-poor conditions) in controlling composition, microstructure, optical, electrical, and electrochemical impedance properties of PLD $Cu_2O$ films were investigated. We found that under suitable deposition conditions both n- and p-type copper oxide thin films can be realized which are discussed below.

## 2. Experimental methods

### 2.1 Growth of PLD films

A simple PLD setup using a UV-ArF Excimer Laser (wavelength: 193 nm, repetition rate: 10 Hz, pulsed width: 20 ns, spot size: ~1 mm$^2$, energy (LP): 25 ± 4 mJ/pulse) was used to deposit copper oxide thin films on amorphous quartz, polycrystalline ITO, and single crystalline NaCl(100) substrates under the following conditions: Base vacuum of the PLD chamber ≤ 10$^{-5}$ Torr; oxygen partial pressure (O$_{2pp}$): 0 – 10 mTorr; target-substrate distance ~ 5 cm; substrate temperature (T$_{sub}$): ranging from non-intentionally heated (RT~ 25 $^0$C) – 400 $^0$C. The substrate temperature was controlled by halogen bulb heater and measured by using a digital thermocouple as described in ref. [25]. The substrate temperature and oxygen partial pressure were varied to allow the film to grow in a stable regime, where no decomposition of the film was observed. Post annealing treatment was also given to some of the as-grown Cu$_2$O films under controlled O$_2$ ambient inside the PLD chamber for comparison purposes.

Prior to film deposition, all substrates were ultrasonically cleaned successively in Toluene Acetone, isopropanol and ultra-pure water (Mili-Q, 18.MΩ.cm) for 15 min followed by a Ar blown dry. All substrates were subjected to an UV-Ozone clean for 20 min immediately before being mounted in the PLD chamber. The target material was commercially available hot pressed ceramic Cu$_2$O (purity~99.95%). The target was ablated for 5 min prior to actual deposition on the substrates.

### 2.2 Characterization of PLD samples

The X-ray Diffraction (XRD) spectra were recorded with Bruker AXS D8 Advance powder X-ray diffractometer using Cu K$_\alpha$(λ=1.5406 Å) radiation. The diffraction patterns were recorded with a step size of ~0.025$^0$ and a time per step of 18 s, the samples were rotated to homogenize the measurements. TEM analyses of samples grown on NaCl(100) were investigated by a JEOL 2010 and a Philips EM 430 as described in ref. [25] where the TEM camera length calibration procedure was also discussed. Raman and Photoluminescence spectra were recorded at room temperature in the backscattering geometry with a Renishaw 2000 confocal spectrometer using λ$_{ext}$ = 514.5 nm Ar-ion laser (P ≤ 5 mW) as an excitation source. The optical transmission and diffuse reflection measurements of films deposited on quartz substrate were made using a UV-VIS-NIR spectrophotometer (Shimadzu UV2600 plus) coupled with an integrating sphere. Sheet resistance was measured by a Keithley 2400 source-measure-unit (SMU) coupled with a custom made collinear 4-point probe. Measurements of highly resistive films (>200 MΩ) were carried out using a precision current source (Keithley 6221 AC/DC) coupled with a Nanovoltmeter (Keithley 2182A). To minimize the non-uniformity nature of the PLD thin film, measurements were performed over three or more different areas of the sample to get a final resistance value. Thicknesses of the PLD films were estimated by combined analyses of FE-SEM (JEOL JSM 6330F) cross-sectional imaging and Variable Angle Spectroscopic Ellipsometry (VASE) (M-2000 U, J.A. Woollam Co.) [37]. Hall mobilities of samples grown on quartz were measured as accurately as possible using a homemade Hall coefficient measurement setup utilizing a commercial 1 Tesla permanent magnet (Magnetsales UK Ltd.) with gold coated electrical contact pads in the van der Pauw sample configuration (see supplementary materials for details). Both Cyclic Voltammetry (CV) and Electrochemical Impedance Spectroscopy (EIS) measurements were performed under dark conditions using a Potentiostat (Autolab, PGSTAT-30) equipped with a frequency analyzer. An AC signal of 20 mV amplitude with frequencies ranging from 50 Hz to 10 kHz was applied at a set of constant bias voltage over a narrow potential window to preserve the stability of the $Cu_2O$ film. The Keithley SMU coupled with a homemade multiprobe workstation was also used to measure I-V curve of the $Cu_2O$-based solid p-n junctions.

## 3. Results and discussion

Four sets of samples were deposited in an attempt to grow high quality single phase copper (I) oxide ($Cu_2O$) at relatively low temperatures with a range of conductivities. In the first two sets of samples, substrate temperature was varied from 25 $^0$C.(RT) to 400 $^0$C with two different constant oxygen partial pressure($O_{2pp}$), namely 3 mTorr and 10 mTorr, in order to find out the optimum process conditions for single phase copper oxide. In the other two sets, the oxygen content inside the PLD chamber was varied ( $0 \leq O_{2pp} \leq 7$ mTorr) at two substrate temperatures ($T_{sub}$), 25 $^0$C and 200 $^0$C. The PLD thin films grown at 25 $^0$C $\leq T_{sub} \leq$ 400 $^0$C with constant $O_{2pp} \approx 10$ mTorr were found to be mainly composed of a mixture of $Cu_2O$ and more O-rich Cu-O phases as evident from the XRD analyses (see Fig. S1a for details). Therefore, the oxygen rich ($O_{2pp} = 10$ mTorr) samples will not be considered further in the discussion below. As a quick guide for the readers: the structural, compositional, and electrical characteristics of samples investigated are summarized in Fig. 1.

The microstructure and surface morphology of Cu (I) oxide thin films are seemingly independent of substrate material as investigated by XRD and FE-SEM where all films are polycrystalline in nature showing grain sizes in the range 50 nm to 150 nm, but with some larger grains and texture on crystalline substrates (see Fig. S1.1 and Fig. S5.1 to S5.4 in supplementary material).

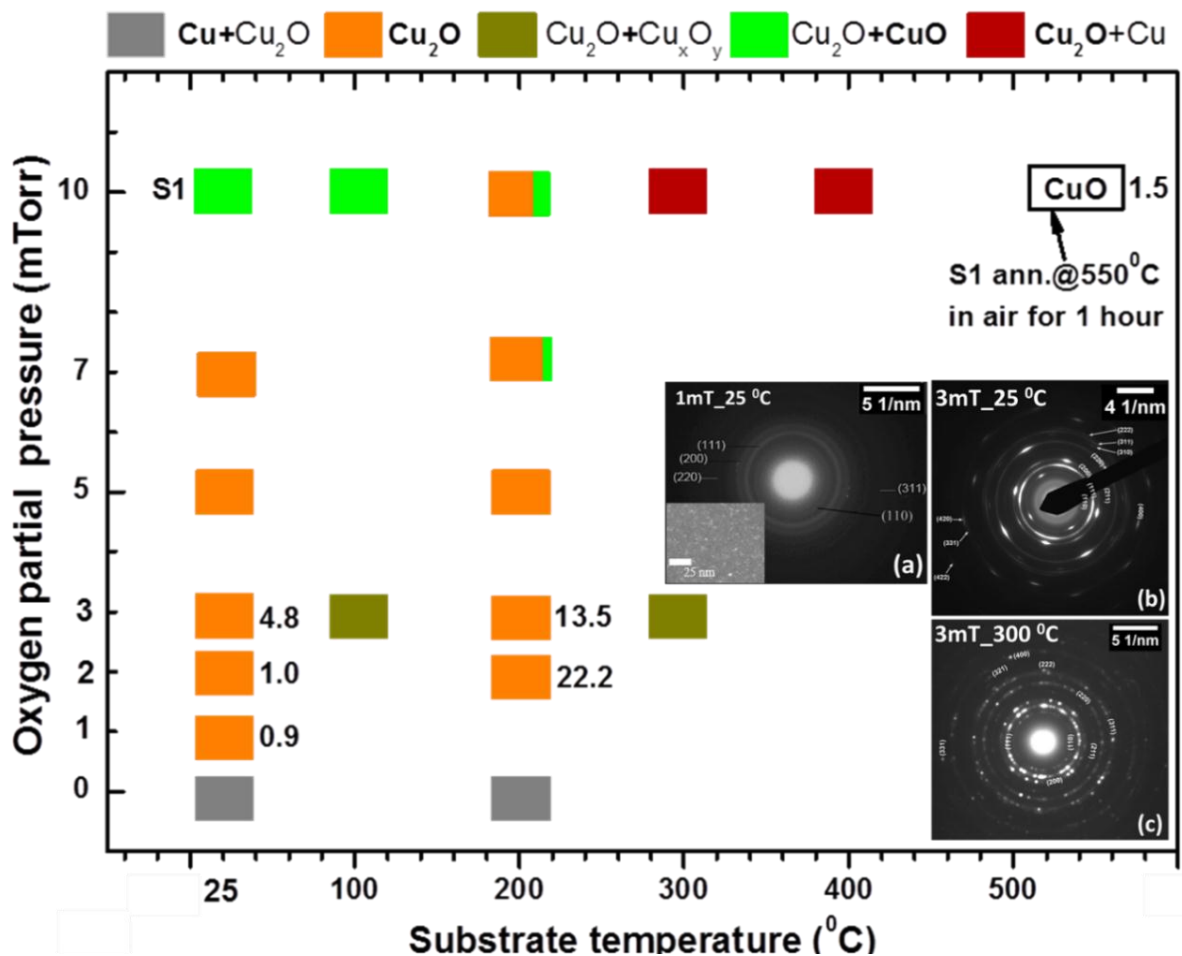

**Fig. 1.** (color online) Deposition diagram of PLD grown copper oxide phases with varying $O_{2pp}$ and $T_{sub}$. The colored squares represent the different phase compositions shown at the top of the graph. The carrier mobility of some as-deposited and annealed S1(Cu(II) oxide: CuO) films are also included in the analysis for comparison purposes. SAED patterns of samples grown on NaCl (100) with (a) $O_{2pp}$ = 1 mTorr_$T_{sub}$ = 25 $^0$C (Inset: bright field image), (b) $O_{2pp}$ = 3 mTorr_$T_{sub}$ = 25 $^0$C, and (c) $O_{2pp}$ = 3 mTorr_$T_{sub}$ = 300 $^0$C are showing single phase Cu(I) oxide with different morphologies.

It can be seen from Fig. 1 (Green, brown, and mixed colored squares) that elevated $T_{sub}$ and high level $O_{2pp} \geq 7$ mTorr produced a mixture of Cu-O phases evident from XRD and Raman analyses (shown in Fig. S1-S2 in supplementary materials). In contrast, substrate temperature 100 $^0$C and 300 $^0$C with fixed $O_{2pp} = 3$ mTorr produce non-stoichiometric Cu (I) oxide ($Cu_2O+Cu_xO_y$) exhibiting cubic SAED patterns similar to phase pure polycrystalline $Cu_2O$ (see also Fig. S1b). The origin of the $Cu_xO_y$ phase - a defect structure of $Cu_2O$ - was discussed in detail in our previous report [25]. It should be noted that depositing films at $T_{sub} = 25$ $^0$C and $T_{sub} = 200$ $^0$C while tuning the $O_{2pp}$ allows one to achieve single phase $Cu_2O$ with varying electrical properties (see orange color squares and numerical value for carrier mobility in Fig. 1). The structural properties of these films are presented in Fig. 2 below.

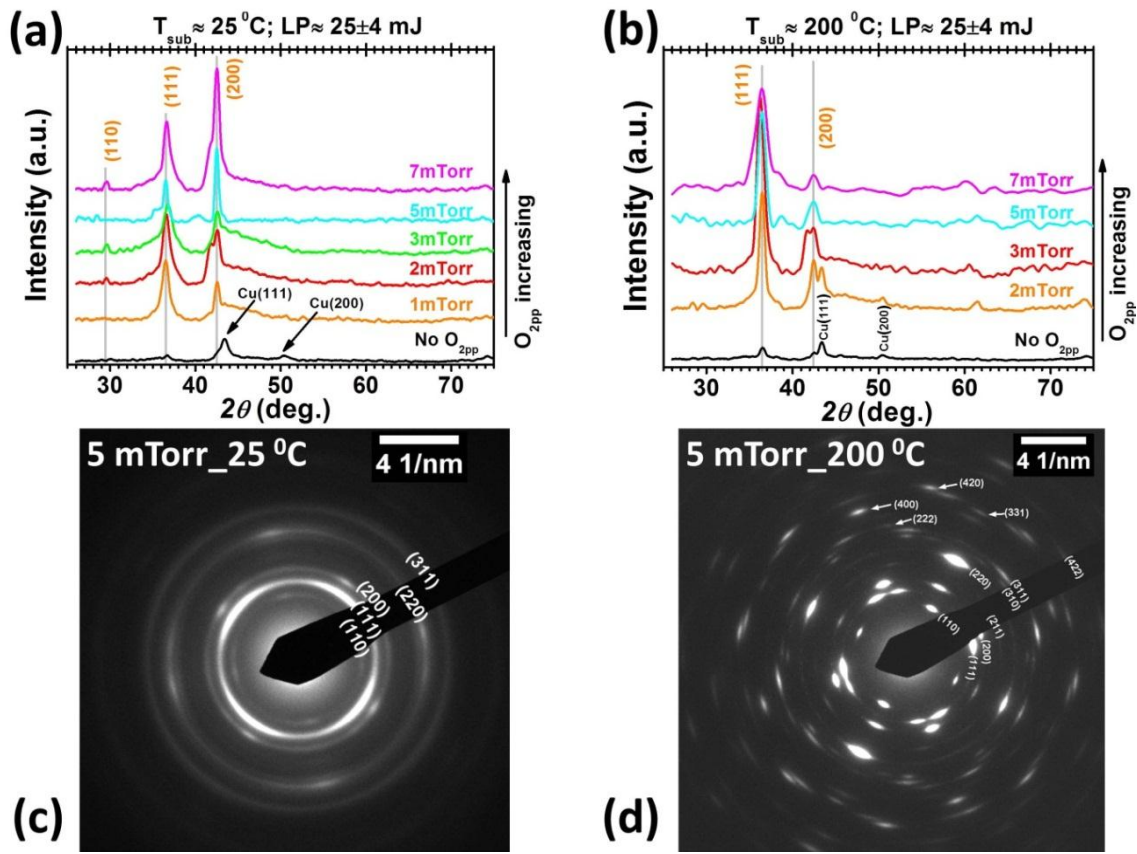

**Fig. 2.** (color online) PLD grown copper oxide thin films deposited at various $O_{2pp}$ at two constant substrate temperatures: XRD patterns of $Cu_2O$ films deposited on quartz glass at (a) $T_{sub} = 25$ $^0$C and (b) $T_{sub} = 200$ $^0$C as a function of $O_{2pp}$. Vertical lines (faint) represent the Bragg's reflection lines for bulk $Cu_2O$. SAED pattern of samples grown on NaCl(100) at $T_{sub} = 25$ $^0$C (c) and $T_{sub} = 200$ $^0$C (d) with a constant $O_{2pp} = 5$ mTorr. The rings/spots are indexed to assist the reader.

Fig. 2a and 2b show the XRD profiles of as-grown films deposited at $T_{sub}$ = 25 $^0$C (RT-grown) and $T_{sub}$ = 200 $^0$C (HT-grown), respectively, on quartz substrates with varying $O_{2pp}$. XRD results revealed that single phase $Cu_2O$, with no contribution from Cu and CuO, could be produced with 1 mTorr ≤ $O_{2pp}$ ≤ 7 mTorr in the case of RT-grown and with 3 mTorr ≤ $O_{2pp}$ ≤ 5 mTorr in the case of HT-grown thin films. In these phase purity limits, both RT- and HT-grown mainly exhibit $Cu_2O(111)$ and $Cu_2O(200)$ Bragg peaks but the former displays an additional $Cu_2O(110)$ peak at 2θ≈29.5$^0$. Metallic Cu inclusions were observed in the RT-grown film with $O_{2pp}$ < 1 mTorr and in the HT-grown film with $O_{2pp}$ ≤ 2 mTorr presumably due to the deficiency of oxygen species in the PLD chamber. However, RT-grown film with $O_{2pp}$ = 1 mTorr show single phase $Cu_2O$ films evident from XRD and SAED pattern (see Fig. 2a and Fig. 1a). Both RT- and HT-grown films with $O_{2pp}$ > 7 mTorr were found to be mixed Cu-O phase (see Fig. S2) with CuO as a major product. Notice also that RT- and HT-grown films approached towards strong {100} and {111} texturing, respectively, with increasing $O_{2pp}$ up to 7 mTorr (see also Fig. S3a). The average crystallite domain size estimated by using Scherrer formula was found to be in the range 5 – 15 nm and found to be following a decreasing trend with increasing $O_{2pp}$ (for details see fig. S3b) most probably due to the inclusion of oxygen rich Cu-O phase at higher $O_{2pp}$ [37]. However, up to $O_{2pp}$ = 5 mTorr, PLD grown thin films are composed of Cu(I) oxide as evident from the SAED patterns shown in Fig. 2c and 2d. The arced SAED patterns, more conspicuous in Fig. 2d, is due to the micro-twinning on ($\bar{1}1\bar{1}$), ($1\bar{1}\bar{1}$), ($1\bar{1}1$) and ($11\bar{1}$) planes while viewed down the [011] zone axis [25].

Room temperature Raman analyses have also been conducted to supplement the XRD results in order to further confirm the phase purity of RT-grown and at HT-grown samples deposited with 2 mTorr ≤ $O_{2pp}$ ≤ 7 mTorr which are presented in Fig. 3 below. All Raman peaks in both groups of samples can be assigned with reported phonon modes of $Cu_2O$ crystal only (solid lines) [25], except for the HT-grown with $O_{2pp}$ = 7 mTorr film which is seen to be a mixture of $Cu_2O$ (solid lines) and CuO (dotted lines) (cf. Fig. 3a and 3b). The presence of CuO phase is confirmed by the additional presence of $A_g$ and $B_g^{(1)}$ modes at ~296 cm$^{-1}$ and ~346 cm$^{-1}$ respectively (see top panel in Fig. 3b). The position of Raman peaks for copper oxide compounds vary considerably in literature, but the vibrational modes for all three copper oxide phases cited in references [6, 25, 42] clearly indicate that none of the Raman signals in figure 3 (up to $O_{2pp}$ ≈ 5 mTorr in the case of HT-grown films) are related to CuO and/or $Cu_4O_3$ phases.

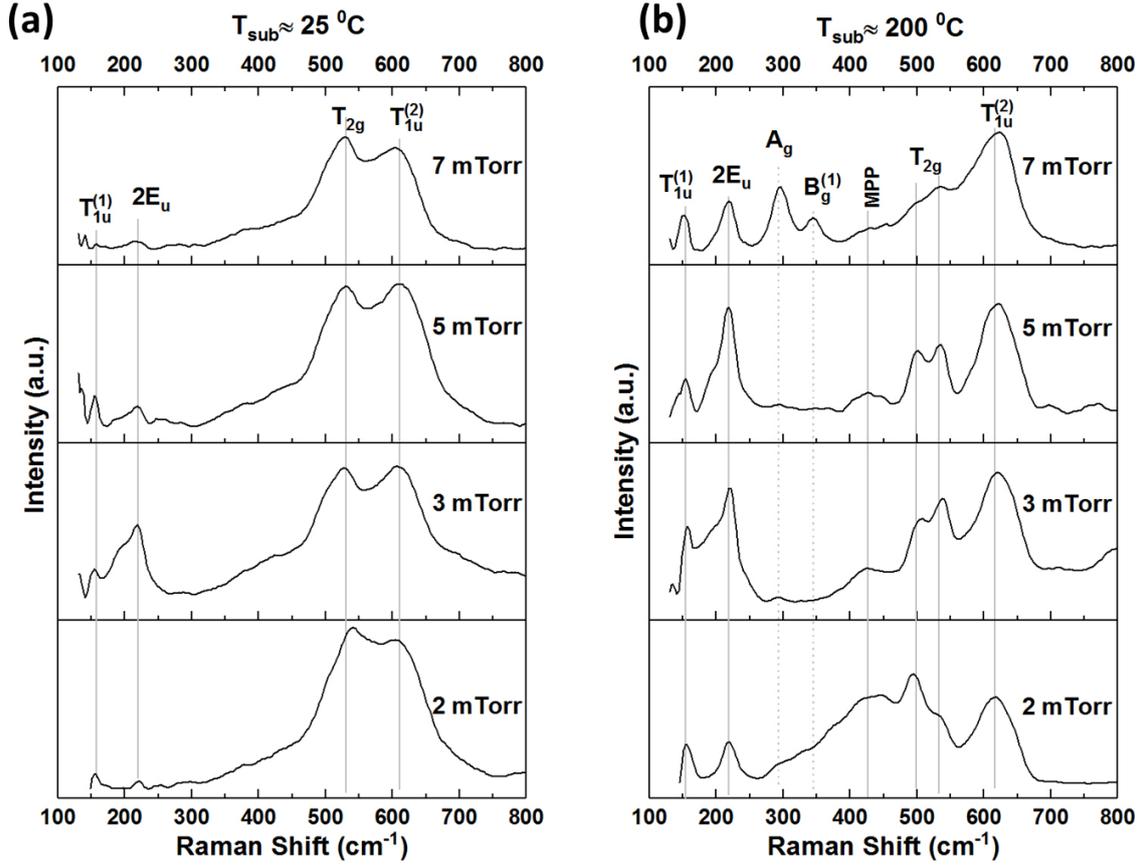

**Fig. 3.** Room temperature Raman spectra of, as-grown copper oxide thin films on quartz glass as a function of $O_{2pp}$ at $T_{sub} \approx 25\ ^0C$ (a), and at $T_{sub} \approx 200\ ^0C$ (b). Vertical lines represent the reference vibrational mode of copper oxide.

The room temperature Photoluminescence (RT-PL) of thin films grown at $T_{sub} \approx 25\ ^0C$(RT) and $T_{sub} \approx 200\ ^0C$(HT) with 2 mTorr $\leq O_{2pp} \leq 5$ mTorr including $Cu_2O$ PLD target are shown in Fig. 4. All luminescence peaks (solid lines) can be attributed to copper oxide phases, consistent with the results reported previously [25]. The exciton-related $(X_0 - \Gamma_{12}^-)$ emission peak is clearly seen in the RT-PL spectrum for all PLD films including the target material suggesting good quality $Cu_2O$ thin films irrespective of $O_{2pp}$ conditions. The luminescence peaks centering at ~760 nm and ~880 nm have been put forward for $Cu_3O_2$ and $[V_{Cu}^- - V_O^+]$ complex respectively [25] . Notice that at low $O_{2pp} = 2 - 3$ mTorr, HT-grown films showed an enhanced $(X_0 - \Gamma_{12}^-)$ peak with additional $[V_{Cu}^- - V_O^+]$ complex peak which is not conspicuous in RT-grown films. These samples also exhibited diminished $V_{cu}$ peak (see Fig. 4b).

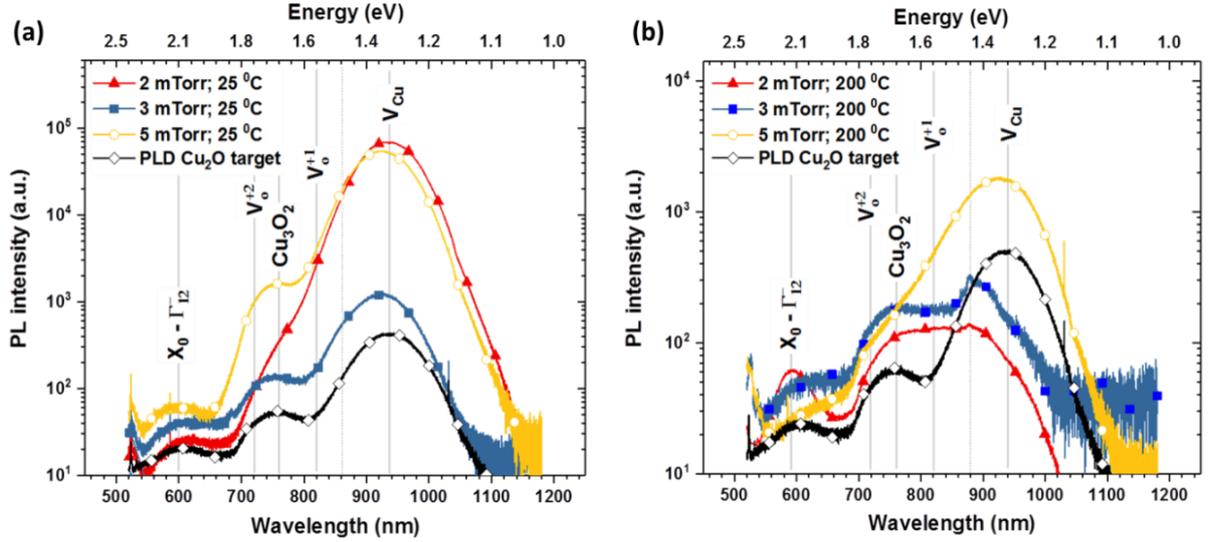

**Fig. 4.** (color online) Room temperature photoluminescence spectra of, as grown copper oxide thin films onto quartz glass as a function of $O_{2pp}$ at $T_{sub} \approx 25\ ^0C$ (a), and at $T_{sub} \approx 200\ ^0C$ (b). Vertical lines represent the luminescence bands of cuprous oxide.

The optical bandgap of the RT- and HT- samples grown with $O_{2pp}$ up to 7 mTorr were estimated by Tauc plot generated from diffuse reflection data using K-M function [43] and shown in Fig. 5. The bandgaps ($E_g$) for RT-grown $Cu_2O$ thin films are found to be gradually and consistently increasing approximately from 1.76 eV to 2.15 eV with increasing $O_{2pp}$ from 1 to 7 mTorr (see Fig. 5a). In contrast, for HT-grown samples, $E_g$ is found to reach a maximum ~2.05 eV for $O_{2pp}$ of 5 mTorr and then to decrease for higher $O_{2pp}$ (see Fig. 5b).

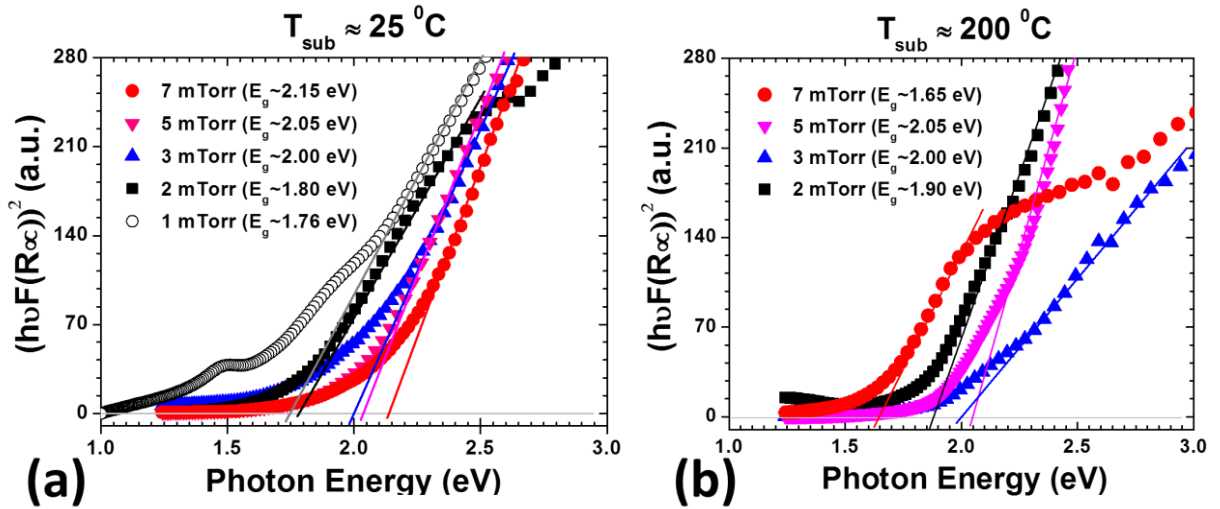

**Fig. 5.** (color online) Comparison of modified Tauc plot of as-grown copper oxide thin films onto quartz glass as a function of $O_{2pp}$ deposited at $T_{sub} \approx 25\ ^0C$ (a), and at $T_{sub} \approx 200\ ^0C$ (b).

The optical band gap thus approaches the bandgap value of bulk $Cu_2O$ ($E_g$ = 2.17 eV), and presumably therefore stoichiometric phase formation, at optimum values of $O_{2pp}$. The lowest measured optical bandgap (~1.65 eV) of HT-grown $O_{2pp}$ = 7 mTorr film suggests an oxygen rich Cu-O phase as also evident from its Raman spectrum (see top panel in Fig. 3b). The RT-grown sample (S1) annealed in air at 550 $^0$C for 1 hour also exhibited a lower optical bandgap ~1.45 eV due to complete conversion to CuO (see Fig. S4 in supplementary material). The lower optical bandgap for films with low $O_{2pp}$, suggest oxygen vacancies ($V_o$) which introduce defect states in the bandgap leading to a lower effective bandgap compared to the stoichiometric $Cu_2O$.

Fig. 6 (below) shows the four point collinear probe measured electrical resistivity of RT- and HT-grown films as a function of $O_{2pp}$. The data points for the films grown without $O_2$ injection in the PLD chamber have also been included in the analyses for the sake of completeness and comparison.

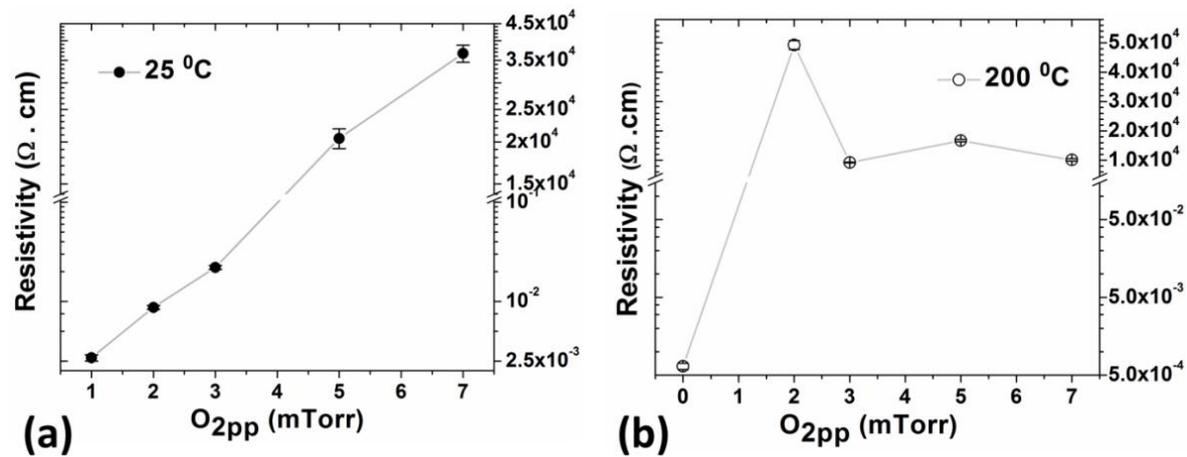

**Fig. 6.** Variation of Electrical resistivity of as-grown copper oxide thin films onto quartz glass as a function of $O_{2pp}$ deposited at $T_{sub}$ ≈ 25 $^0$C (a) and $T_{sub}$ ≈ 200 $^0$C (b).

As can be seen from the Fig. 6a, the resistivities of RT-grown films are found to be monotonically increasing with increasing $O_{2pp}$. At 1 mTorr ≤ $O_{2pp}$ ≤ 3 mTorr PLD ambient their resistivities were estimated to be roughly 3 – 24 mΩ.cm which is ~3 orders of magnitude lower than the reported resistivity (40 – 60 Ω.cm) for $Cu_2O$ films grown by PLD above $T_{sub}$ ≥ 600 $^0$C [44]. On the other hand, the resistivities of RT-films with 5 mTorr ≤ $O_{2pp}$ ≤ 7 mTorr PLD ambient are quite high, roughly in the range 20 – 38 kΩ.cm (see Fig. 6a) due to more stoichiometric $Cu_2O$ phase formed in the oxygen rich conditions. In contrast, the resistivity of HT-grown films without injecting $O_2$ into the PLD chamber was estimated to be below 7 mΩ.cm (see Fig. 6b). The HT-grown film deposited using $O_{2pp}$ = 2 mTorr exhibited the highest resistivity of ~49 kΩ.cm among all samples, due to inclusion of metallic Cu into the $Cu_2O$ matrix which compensates available holes leading to insulating thin films. Since, the PLD chamber provides highly non-equilibrium growth environment to the energetic ablated particles and therefore, the probability to exist metallic Cu as ionized defect states ($Cu_i^+$) is very high in oxygen poor heated substrate surface. These energetic $Cu_i^+$ defects compensate available holes leading to insulating thin films compared to the films grown in oxygen rich conditions. Similar observation was reported by X. Liu et al. for PLD grown $Cu_2O$ ($T_{sub}$ = 600 $^0C$ on MgO(110)) with varying $O_{2pp}$ [26]. In fact, the resistivities of 2 mTorr ≤ $O_{2pp}$ ≤ 7 mTorr HT-grown films are roughly six orders of magnitude higher than that of the film grown without oxygen. These results are also suggesting that oxygen rich PLD ambient leads to more stoichiometric copper oxide, thereby resulting in more insulating thin films due to the lack of charge carrier creating defect (vacancy type) concentrations.

Hall coefficient measurements and Mott-Schottky plots (see Fig. 7a) constructed from electrochemical impedance spectra [9, 33] were also carried out for samples grown with identical deposition PLD conditions on quartz and ITO substrates respectively (see Fig. S7 in supplementary material). The results together with optical bandgap values are summarized in Table 1. As can be seen, samples grown with $O_{2pp}$ = 2 – 3 mTorr, showed a p-type conductivity with appreciable Hall mobility ($\mu_H$) ~ 4.78 ± 0.01 $cm^2$ $V^{-1}s^{-1}$ and ~ 22.20 ± 0.01 $cm^2$ $V^{-1}s^{-1}$ respectively in RT- and HT-grown PLD $Cu_2O$ films. The higher mobility in HT-grown PLD films could be attributed to the better crystalline and optoelectronic quality evident from XRD, SAED, Raman, and RT-PL analyses. However, the lower carrier concentrations (~ 6×10$^{12-13}$ $cm^{-3}$) of HT-grown films may be attributed to the hole killing [$V_{Cu}^-$ - $V_O^+$] defect complex seen in PL spectra (see Fig. 4b).

**Table 1**

Optical and Electrical properties of single phase $Cu_2O$ and $CuO$ thin films deposited with low $O_{2pp}$ content during growth.

| Growth condition ($T_{sub}\_O_{2pp}$) | Band gap (eV) ± 0.01 | Thickness (nm) (VASE/ SEM) | Resistivity ($\Omega.cm$) | Carrier density ($cm^{-3}$) [Hall effect meas.] | Hall Mobility ($cm^2/V.s$) ± 0.01 | Carrier density ($cm^{-3}$) [EIS meas.] |
|---|---|---|---|---|---|---|
| 25_1 | 1.76 | 556±21 | $2.12\times10^{-3}$ | $-3.47\times10^{21}$ | 0.85 | $-(1.24\pm.03)\times10^{21}$ |
| 25_2 | 1.80 | 562±4 | $8.71\times10^{-3}$ | $6.97\times10^{20}$ | 1.03 | - |
| 25_3 | 2.00 | 615±25 | $24.00\times10^{-3}$ | $5.45\times10^{19}$ | 4.78 | $(1.39\pm.01)\times10^{20}$ |
| 200_2 | 1.90 | 510±7 | $49.28\times10^{3}$ | $5.71\times10^{12}$ | 22.21 | - |
| 200_3 | 2.00 | 585±7 | $9.25\times10^{3}$ | $5.02\times10^{13}$ | 13.45 | - |
| Ann. S1(CuO) | 1.50 | Assumed ~500 | $12.2 \pm 0.7$ | $-3.42\times10^{17}$ | 1.50 | - |

The acceptor concentration, $N_a \sim 6\times10^{12-13}$ $cm^{-3}$ and mobility, $\mu_H \sim 13\text{-}22$ $cm^2$ $V^{-1}s^{-1}$ estimated for $T_{sub} \approx 200$ $^0C$(HT) samples are consistent with previous results [6]. The high mobility indicates that PLD films grown at 200 $^0C$ possess reasonably good electrical quality, however, a carrier concentration as high as $\sim 10^{14} - 10^{16}$ $cm^{-3}$ is desirable for potential $Cu_2O$ based device applications [27]. Therefore, for some specific device applications [3, 26, 45] RT-grown PLD $Cu_2O$ may be utilized.

Notice that carrier density in RT-grown samples is exceptionally high compared to that of HT-grown samples. For example, with $O_{2pp} = 2$ mTorr, carrier density $\sim(5.71\pm0.18)\times10^{12}$ $cm^{-3}$ obtained for HT-grown sample is at least eight orders of magnitude lower than the carrier density $\sim(6.97\pm0.01)\times10^{20}$ $cm^{-3}$ for RT-grown samples. With increasing $O_{2pp}$ content slightly to 3 mTorr for RT-grown sample, it was obtained as $\sim(5.45\pm0.23)\times10^{19}$ $cm^{-3}$ which is still roughly six orders of magnitude higher than its HT-grown counterpart.

The effective density of states (D(E$_f$)) for these samples are in the range (D(E$_f$)) ~ 1.35 – 5.67×10$^{21}$ (eV)$^{-1}$ cm$^{-3}$ estimated from EIS data by a similar approach presented in ref. [46] using the equation : $D(E_f) = (C_{sc})^2 / [\varepsilon_0 \varepsilon . e^2]$, where, $C_{sc}$ is the space charge capacitance, $\varepsilon_0$ is permittivity of free space (8.854×10$^{-12}$ F.m$^{-1}$), $\varepsilon$ is the relative dielectric constant (~6.6 for Cu$_2$O [33]) and $e$ is the electronic charge (see supplemental material for detail of the calculations). The high level of N$_a$, is consistent with low resistivity (few mΩ.cm), and low Hall mobilities (μ$_H$) ~1.03 – 4.78 cm$^2$V$^{-1}$s$^{-1}$ and could be attributed to the scattering of holes at the grain boundaries of these RT-grown nanocrystalline films.

In contrast, the samples grown at 1 mTorr_ 25 $^0$C(RT) exhibited an n-type conductivity with high level of donor concentrations (N$_d$) ~ -3.47×10$^{21}$ cm$^{-3}$ (μ$_H$ ~0.85 ± 0.01 cm$^2$ V$^{-1}$s$^{-1}$) and ~-1.24×10$^{21}$ cm$^{-3}$ obtained from Hall coefficient measurements and M-S analyses respectively. The effective density of states (D(E$_f$)) for this sample is 1.05×10$^{20}$ (eV)$^{-1}$ cm$^{-3}$ and is slightly lower when compared to p-type RT-grown samples. However, such high level of carrier concentrations may contribute to its exceptionally low resistivity of ~3 mΩ.cm. The optical bandgap (~1.76 eV) of this n-type sample suggests that the donor level may be shallow. In the literature, the Fermi level of donor and acceptor levels for intrinsic Cu$_2$O is rarely reported ([9],[22] and refs. therein). An approximate band diagram of a PLD-grown Cu$_2$O thin film electrode is shown in Fig. 7b using electrochemical M-S plot and UV-VIS-NIR optical data to demonstrate the quasi Fermi level energy of $E_f^e$ (~0.28 eV below CBM) and $E_f^h$ (~0.13 eV above VBM) for donor and acceptor states respectively. The Fermi level is estimated from Flat band potentials -0.72 V *vs* RHE (Fermi energy with respect to vacuum, E$_f$ = -3.73 eV) and +0.91 V *vs* RHE ( E$_f$ = - 5.37 eV) respectively for n- and p-type Cu$_2$O/ITO electrodes using a similar approach as described in ref. [9]. The distinct positive- and negative- slope seen in M-S plots in Fig. 7a confirms the n- and p-type conductivity of Cu$_2$O/ITO electrodes and as seen in Fig. 7b the donor level is obviously located in more positive potentials than the acceptor level for Cu$_2$O which corroborate the reported results [9, 24] . The estimated $E_f^e$ ~0.28 eV below the conduction band minimum (CBM) is considerably shallower than the results of Garuthara et al. (~0.38 eV) [47] and $E_f^h$ ~0.13 eV above the valence band maximum (VBM) which is slightly small compared to ~0.22 eV for V$_{cu}$ and ~0.47 eV for V$_{cu,split}$ [22] .

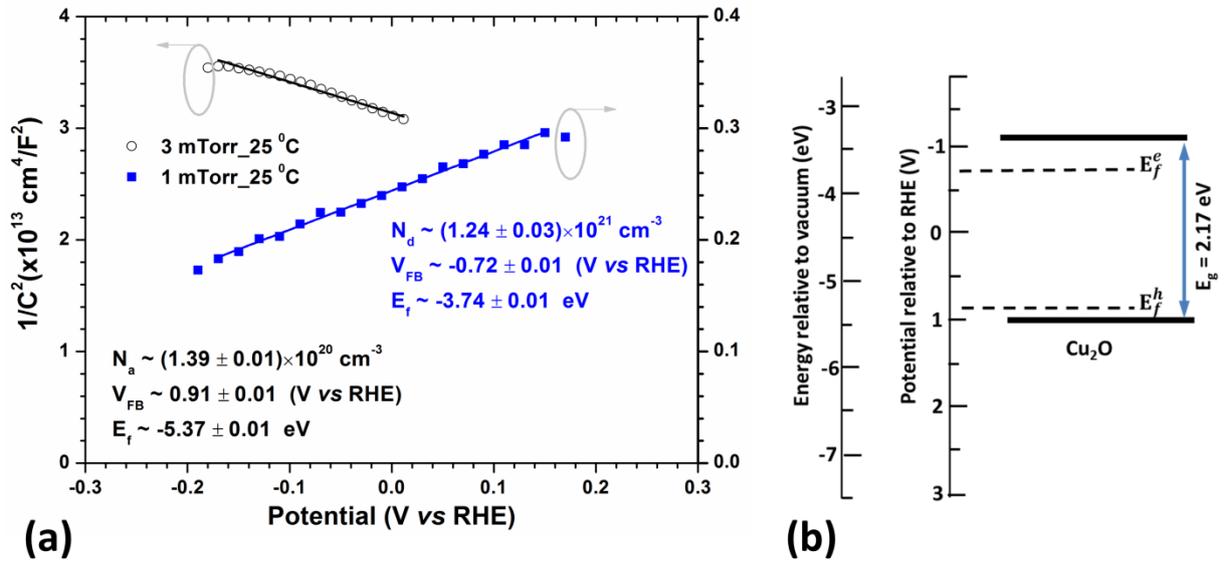

**Fig. 7.** (a) Mott-Schottky plots of $Cu_2O$ thin films grow on ITO using 1 mTorr_25 $^0$C and 3 mTorr_25 $^0$C deposition conditions showing n-type and p-type conductivity respectively. The estimated carrier concentration, Flat band potential ($V_{FB}$) and quasi Fermi level energy ($E_f^x$; where $x = e$ or $h$) for donor and acceptor states are also shown in the inset. (b) An approximate band diagram of $Cu_2O$ grown at low $O_{2pp}$ and $T_{sub}$ by PLD.

The low resistivity ($\rho \sim 3$ mΩ.cm) of n-type $Cu_2O$ thin films grown by PLD is yet at least ~3 orders of magnitude above the resistivity of thermally evaporated metallic Cu (3.5 µΩ.cm) and Cu+$Cu_2O$ (27 µΩ.cm) thin films reported by Figueiredo et al. [30]. Additionally, samples grown at 1 mTorr_ 25 $^0$C(RT) are single phase $Cu_2O$, this is evident from both XRD (Fig. 2a) and TEM (Fig. 1a) analyses and they are the most oxygen deficient $Cu_2O$ among the samples deposited in this study. Therefore, the origin of n-type conductivity in these PLD films is presumably due to the electron generating predominant oxygen vacancy ($V_o$) at the vicinity of Cu/$Cu_2O$ phase boundary [23, 39]. The single phase n-type $Cu_2O$ deposited by electrodeposition with $N_d$ as high as ~$10^{20}$ cm$^{-3}$ [15] has repeatedly been appeared in the literature, but single phase n-type $Cu_2O$ deposited by PLD is rarely reported. Recently, Xu et al. [11] reported PLD grown phase pure $Cu_2O$ using $O_{2pp} = 0.09$ Pa at $T_{sub} = 600$ $^0$C with p-type conductivity and they were able to convert them into n-type $Cu_2O$ via post-deposition $N_2$ plasma treatment. However, in our study the slight increase of $O_{2pp}$ from 1 to 2 mTorr or more in the PLD chamber yielded p-type $Cu_2O$ films suggesting that n-type $Cu_2O$ can only be stabilized in a narrow range of $O_{2pp} \leq 1$mTorr. Our findings for RT-grown samples suggest that critical regulation of low level $O_{2pp}$ helps achieving single phase $Cu_2O$ with tunable optical band gap ($E_g$ = 1.76 – 2.15 eV) and type of conductivity with appreciable Hall mobilities ($\mu_H$ ~0.85 – 4.78 $cm^2$ $V^{-1}s^{-1}$). Compared to results reported by Xu et al. this is important for synthesizing diverse optoelectronic device with low thermal budget [3, 45] including p-n $Cu_2O$ based homojunctions [7, 9, 12, 14, 24].

It is conspicuous that estimated acceptor concentrations in RT-grown p-Type $Cu_2O$ samples are also high ($N_a$ ~ $6\times10^{19-20}$ $cm^{-3}$) in our observations (see Table 1 and Fig. 7). This is consistent with other recent reports. Theoretical studies reported by Raebiger et al. [16] suggest that the dominant defect $V_{Cu}$ can be above $10^{21}$ $cm^{-3}$ in concentrations, both in Cu-rich(oxygen poor) and Cu-poor(oxygen rich) growth conditions, to produce a p-type conductivity. Recent experimental works conducted by Zhang et al. [48] reported hole concentrations in the range $5.81\times10^{18}$ – $2.17\times10^{21}$ $cm^{-3}$ for sputtered $Cu_2O$ thin films grown at room temperatures. Chen et al. [49] also reported that hole concentrations could be as high as $1.5\times10^{21}$ $cm^{-3}$.

The PLD sample (S1) annealed at 550 $^0$C in air for 1 hour is single phase CuO and showed n-type conductivity with carrier concentration ~ $3.47\times10^{17}$ $cm^{-3}$ and mobility ~1.5 $cm^2$ $V^{-1}s^{-1}$ and is consistent with reported results [31]. The n-type conductivity of CuO can be understood as follows. PLD grown $Cu_2O$ thin films with high $O_{2pp}$ are copper deficient p-type semiconductors where the type of conductivity arises from the copper vacancies can be described by following equation:

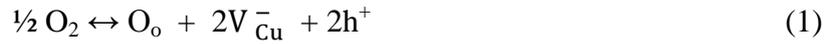

$$\tfrac{1}{2} O_2 \leftrightarrow O_o + 2V_{Cu}^{-} + 2h^+ \qquad (1)$$

where, $O_O$ - oxygen in regular positions of the lattice, $V_{Cu}^{-}$ - metal vacancies, and $h^+$- holes. Due to annealing of p-type $Cu_2O$ in air, more and more oxygen species are escaped from the $Cu_2O$ matrix and generate more $Cu^{2+}$ at the expense of $Cu^{1+}$ in the films (this recrystallizes to a CuO phase) leading to the creation of an electron-generating oxygen vacancy ($V_o^{+2}$) according to the following equation:

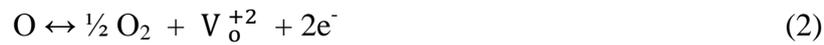

$$O \leftrightarrow \tfrac{1}{2} O_2 + V_o^{+2} + 2e^- \qquad (2)$$

where, $V_o^{+2}$ - oxygen vacancies, and $e^-$ - electrons. As shown in equation (2), evolution of oxygen is accompanied by the formation of oxygen vacancies (with effective positive charge) and electrons that determine n-type of conductivity.

As a proof-of-concept, solid p-n junctions were fabricated with FTO/n-ZnO/p-$Cu_2O$/Au and p-Si(111)/n-$Cu_2O$ structures. Their stable current-voltage characteristic curves suggest that p-n junctions were formed successfully, albeit with poor photovoltaic performance. These results are summarized and discussed in the supplementary materials (see Fig. S12 - S14).

## 4. Conclusions

Single phase n- and p-type $Cu_2O$ thin films were grown on quartz glass and other substrates (ITO, NaCl(100), and p-Si(111)) by a PLD technique and the influence of substrate temperatures and oxygen pressure ($O_{2pp}$) on the films properties were explored systematically. Thin films grown at 25 $^0$C(RT) and 200 $^0$C(HT) substrate temperatures are single phase $Cu_2O$ with (200) and (111) texture respectively and their texturing is found to be increasing with the increasing oxygen content in the 2 mTorr ≤ $O_{2pp}$ ≤ 5 mTorr PLD ambient. The optical bandgaps and electrical resistivities of RT-grown $Cu_2O$ thin films were found to be tunable in the range 1.76 eV – 2.15 eV and 3 mΩ.cm – 38 kΩ.cm respectively and both properties were found to be consistently increasing from low to higher values with increasing $O_{2pp}$ due to more stoichiometric Cu(I) phase formation. Critical regulations of low level $O_{2pp}$ during the growth process helps achieving both n- and p-type $Cu_2O$ with appreciable Hall mobilities ($\mu_H$ ~0.85 – 4.78 $cm^2 V^{-1}s^{-1}$). The as-grown p- and n-type $Cu_2O$ showed promising rectification in solid junctions with n-ZnO and p-Si electrodes respectively. The findings reported here suggest that critical regulations of PLD growth conditions help achieve single phase $Cu_2O$ growth with tunable electronic properties desirable for diverse optoelectronic applications.

**Acknowledgments**

S.F.U. Farhad acknowledges the financial help through BANGABANDHU fellowship, Ministry of Science and Technology, Government of Bangladesh to conduct this research. The authors are indebted to Professor Mike Ashfold FRS for allowing his PLD setup located at the Diamond laboratory, School of Chemistry, University of Bristol, UK during the Ph.D. studies. Special thanks are also due to Professor Walther Schawrzer, School of Physics, University of Bristol, UK for lending his Kiethley SMU2400 instrument in building a custom-made 4-point collinear probe resistivity measurement setup. S.F.U Farhad also acknowledges the PV characterization support of the Energy Conversion and Storage Research (ECSR) section, Industrial Physics Division, BCSIR Labs, Dhaka, Bangladesh.

**Conflict of interest:** We declare that we have no conflict of interest.

**Author Contributions**: S.F.U. Farhad and D.Cherns conceived the idea . D.Cherns supervised the work. S.F.U. Farhad did the experimental works and wrote the manuscript. J.Smith and N. Fox helped with the deposition setup and Hall coefficient measurements. D. Fermín advised on the electrochemical measurements and analyses. All authors contributed to editing and reviewing the final manuscript.

**References**

[1] H. Kim, S. Bae, D. Jeon, J. Ryu, Fully solution-processable $Cu_2O$–$BiVO_4$ photoelectrochemical cells for bias-free solar water splitting, Green Chemistry, 20 (2018) 3732-3742. doi:10.1039/c8gc00681d

[2] Y.S. Zhi, P.G. Li, P.C. Wang, D.Y. Guo, Y.H. An, Z.P. Wu, X.L. Chu, J.Q. Shen, W.H. Tang, C.R. Li, Reversible transition between bipolar and unipolar resistive switching in $Cu_2O$/$Ga_2O_3$ binary oxide stacked layer, AIP Advances, 6 (2016) 015215. doi:10.1063/1.4941061

[3] D. Muñoz-Rojas, M. Jordan, C. Yeoh, A.T. Marin, A. Kursumovic, L.A. Dunlop, D.C. Iza, A. Chen, H. Wang, J.L. MacManus Driscoll, Growth of ∼5 $cm^2V^{-1}s^{-1}$ mobility, p-type Copper (I) oxide ($Cu_2O$) films by fast atmospheric atomic layer deposition (AALD) at 225°C and below, AIP Advances, 2 (2012) 042179. doi:10.1063/1.4771681

[4] H. Zhang, Q. Zhu, Y. Zhang, Y. Wang, L. Zhao, B. Yu, One-Pot Synthesis and Hierarchical Assembly of Hollow $Cu_2O$ Microspheres with Nanocrystals-Composed Porous Multishell and Their Gas-Sensing Properties, Advanced Functional Materials, 17 (2007) 2766-2771. doi:10.1002/adfm.200601146

[5] K. Chen, S. Song, D. Xue, Faceted $Cu_2O$ structures with enhanced Li-ion battery anode performances, CrystEngComm, 17 (2015) 2110-2117. doi:10.1039/c4ce02340d

[6] B.K. Meyer, A. Polity, D. Reppin, M. Becker, P. Hering, P.J. Klar, T. Sander, C. Reindl, J. Benz, M. Eickhoff, C. Heiliger, M. Heinemann, J. Bläsing, A. Krost, S. Shokovets, C. Müller, C. Ronning, Binary copper oxide semiconductors: From materials towards devices, physica status solidi (b), 249 (2012) 1487-1509. doi:10.1002/pssb.201248128

[7] L.C. Olsen, F.W. Addis, W. Miller, Experimental and theoretical studies of $Cu_2O$ solar cells, Solar Cells, 7 (1982) 247-279. doi:10.1016/0379-6787(82)90050-3

[8] A.E. Rakhshani, Preparation, characteristics and photovoltaic properties of cuprous oxide—a review, Solid-State Electronics, 29 (1986) 7-17. doi:10.1016/0038-1101(86)90191-7

[9] C.M. McShane, K.S. Choi, Junction studies on electrochemically fabricated p-n $Cu_2O$ homojunction solar cells for efficiency enhancement, Physical chemistry chemical physics : PCCP, 14 (2012) 6112-6118. doi:10.1039/c2cp40502d

[10] L. Xiong, S. Huang, X. Yang, M. Qiu, Z. Chen, Y. Yu, p-Type and n-type $Cu_2O$ semiconductor thin films: Controllable preparation by simple solvothermal method and photoelectrochemical properties, Electrochimica Acta, 56 (2011) 2735-2739. doi:10.1016/j.electacta.2010.12.054

[11] M. Xu, X. Liu, W. Xu, H. Xu, X. Hao, X. Feng, Low resistivity phase-pure n-type $Cu_2O$ films realized via post-deposition nitrogen plasma treatment, Journal of Alloys and Compounds, 769 (2018) 484-489. doi:10.1016/j.jallcom.2018.08.048

[12] R.P. Wijesundera, L.K.A.D.D.S. Gunawardhana, W. Siripala, Electrodeposited $Cu_2O$ homojunction solar cells: Fabrication of a cell of high short circuit photocurrent, Solar Energy Materials and Solar Cells, 157 (2016) 881-886. doi:10.1016/j.solmat.2016.07.005

[13] J. Han, J. Chang, R. Wei, X. Ning, J. Li, Z. Li, H. Guo, Y. Yang, Mechanistic investigation on tuning the conductivity type of cuprous oxide ($Cu_2O$) thin films via deposition potential, International Journal of Hydrogen Energy, 43 (2018) 13764-13777. doi:10.1016/j.ijhydene.2018.02.121

[14] K. Han, M. Tao, Electrochemically deposited p–n homojunction cuprous oxide solar cells, Solar Energy Materials and Solar Cells, 93 (2009) 153-157. doi:10.1016/j.solmat.2008.09.023

[15] P. Wang, H. Wu, Y. Tang, R. Amal, Y.H. Ng, Electrodeposited $Cu_2O$ as Photoelectrodes with Controllable Conductivity Type for Solar Energy Conversion, The Journal of Physical Chemistry C, 119 (2015) 26275-26282. doi:10.1021/acs.jpcc.5b07276

[16] H. Raebiger, S. Lany, A. Zunger, Origins of the p-type nature and cation deficiency in $Cu_2O$ and related materials, Physical Review B, 76 (2007). doi:10.1103/PhysRevB.76.045209

[17] L.Y. Isseroff, E.A. Carter, Electronic Structure of Pure and Doped Cuprous Oxide with Copper Vacancies: Suppression of Trap States, Chemistry of Materials, 25 (2013) 253-265. doi:10.1021/cm3040278

[18] K.P. Musselman, A. Marin, A. Wisnet, C. Scheu, J.L. MacManus-Driscoll, L. Schmidt-Mende, A Novel Buffering Technique for Aqueous Processing of Zinc Oxide Nanostructures and Interfaces, and Corresponding Improvement of Electrodeposited ZnO- $Cu_2O$ Photovoltaics, Advanced Functional Materials, 21 (2011) 573-582. doi:10.1002/adfm.201001956

[19] S.S. Wilson, J.P. Bosco, Y. Tolstova, D.O. Scanlon, G.W. Watson, H.A. Atwater, Interface stoichiometry control to improve device voltage and modify band alignment in ZnO/ $Cu_2O$ heterojunction solar cells, Energy Environ. Sci., 7 (2014) 3606-3610. doi:10.1039/c4ee01956c

[20] M. Pavan, S. Rühle, A. Ginsburg, D.A. Keller, H.-N. Barad, P.M. Sberna, D. Nunes, R. Martins, A.Y. Anderson, A. Zaban, E. Fortunato, $TiO_2$/ $Cu_2O$ all-oxide heterojunction solar cells produced by spray pyrolysis, Solar Energy Materials and Solar Cells, 132 (2015) 549-556. doi:10.1016/j.solmat.2014.10.005

[21] K. Kardarian, D. Nunes, P. Maria Sberna, A. Ginsburg, D.A. Keller, J. Vaz Pinto, J. Deuermeier, A.Y. Anderson, A. Zaban, R. Martins, E. Fortunato, Effect of Mg doping on $Cu_2O$ thin films and their behavior on the $TiO_2/Cu_2O$ heterojunction solar cells, Solar Energy Materials and Solar Cells, 147 (2016) 27-36. doi:10.1016/j.solmat.2015.11.041

[22] D.O. Scanlon, G.W. Watson, Undoped n-Type $Cu_2O$: Fact or Fiction?, The Journal of Physical Chemistry Letters, 1 (2010) 2582-2585. doi:10.1021/jz100962n

[23] L. Frazer, K.B. Chang, R.D. Schaller, K.R. Poeppelmeier, J.B. Ketterson, Vacancy relaxation in cuprous oxide ($Cu_{2-x}O_{1-y}$), Journal of Luminescence, 183 (2017) 281-290. doi:10.1016/j.jlumin.2016.11.011

[24] L.C.-K. Liau, Y.-C. Lin, Y.-J. Peng, Fabrication Pathways of p–n $Cu_2O$ Homojunction Films by Electrochemical Deposition Processing, The Journal of Physical Chemistry C, 117 (2013) 26426-26431. doi:10.1021/jp405715c

[25] S.F.U. Farhad, R.F. Webster, D. Cherns, Electron microscopy and diffraction studies of pulsed laser deposited cuprous oxide thin films grown at low substrate temperatures, Materialia, 3 (2018) 230 - 238. doi:10.1016/j.mtla.2018.08.032

[26] X. Liu, M. Xu, X. Zhang, W. Wang, X. Feng, A. Song, Pulsed-laser-deposited, single-crystalline $Cu_2O$ films with low resistivity achieved through manipulating the oxygen pressure, Applied Surface Science, 435 (2018) 305-311. doi:10.1016/j.apsusc.2017.11.119

[27] S.H. Wee, P.S. Huang, J.K. Lee, A. Goyal, Heteroepitaxial $Cu_2O$ thin film solar cell on metallic substrates, Scientific reports, 5 (2015) 16272. doi:10.1038/srep16272

[28] S. Ishizuka, K. Akimoto, Control of the growth orientation and electrical properties of polycrystalline $Cu_2O$ thin films by group-IV elements doping, Applied Physics Letters, 85 (2004) 4920. doi:10.1063/1.1827352

[29] S. Nandy, R. Thapa, M. Kumar, T. Som, N. Bundaleski, O.M.N.D. Teodoro, R. Martins, E. Fortunato, Efficient Field Emission from Vertically Aligned $Cu_2O_{1-\delta}$(111) Nanostructure Influenced by Oxygen Vacancy, Advanced Functional Materials, 25 (2015) 947-956. doi:10.1002/adfm.201402910

[30] V. Figueiredo, E. Elangovan, G. Gonçalves, P. Barquinha, L. Pereira, N. Franco, E. Alves, R. Martins, E. Fortunato, Effect of post-annealing on the properties of copper oxide thin films obtained from the oxidation of evaporated metallic copper, Applied Surface Science, 254 (2008) 3949-3954. doi:10.1016/j.apsusc.2007.12.019

[31] X. Hu, F. Gao, Y. Xiang, H. Wu, X. Zheng, J. Jiang, J. Li, H. Yang, S. Liu, Influence of oxygen pressure on the structural and electrical properties of CuO thin films prepared by pulsed laser deposition, Materials Letters, 176 (2016) 282-284. doi:10.1016/j.matlet.2016.04.055

[32] N. Gupta, R. Singh, F. Wu, J. Narayan, C. McMillen, G.F. Alapatt, K.F. Poole, S.-J. Hwu, D. Sulejmanovic, M. Young, G. Teeter, H.S. Ullal, Deposition and characterization of nanostructured $Cu_2O$ thin-film for potential photovoltaic applications, Journal of Materials Research, 28 (2013) 1740-1746. doi:10.1557/jmr.2013.150

[33] A. Paracchino, J.C. Brauer, J.-E. Moser, E. Thimsen, M. Graetzel, Synthesis and Characterization of High-Photoactivity Electrodeposited $Cu_2O$ Solar Absorber by Photoelectrochemistry and Ultrafast Spectroscopy, The Journal of Physical Chemistry C, 116 (2012) 7341-7350. doi:10.1021/jp301176y

[34] S.F.U. Farhad, M.M. Hossain, N.I. Tanvir, S. Islam, Texture and Bandgap Tuning of Phase Pure $Cu_2O$ Thin Films Grown By a Simple Potentiostatic Electrodeposition Technique, ECS Meeting Abstracts, MA2020-01 (2020) 1212-1212. doi:10.1149/ma2020-01191212mtgabs

[35] S.F.U. Farhad, M.A. Hossain, N.I. Tanvir, R. Akter, M.A.M. Patwary, M. Shahjahan, M.A. Rahman, Structural, optical, electrical, and photoelectrochemical properties of cuprous oxide thin films grown by modified SILAR method, Materials Science in Semiconductor Processing, 95 (2019) 68-75. doi:10.1016/j.mssp.2019.02.014

[36] M.Y. Ghotbi, Z. Rahmati, Nanostructured copper and copper oxide thin films fabricated by hydrothermal treatment of copper hydroxide nitrate, Materials & Design, 85 (2015) 719-723. doi:10.1016/j.matdes.2015.07.081

[37] S.F.U. Farhad, Copper Oxide Thin Films grown by Pulsed Laser Deposition for Photovoltaic Applications, in: School of Physics, University of Bristol, UK, January 2016, pp. 222. https://ethos.bl.uk/OrderDetails.do?uin=uk.bl.ethos.691178

[38] K. Wang, W. Gao, H.W. Zheng, F.Z. Li, M.S. Zhu, G. Yang, G.T. Yue, Y.K. Liu, R.K. Zheng, Heteroepitaxial growth of $Cu_2O$ films on Nb-$SrTiO_3$ substrates and their photovoltaic properties, Ceramics International, 43 (2017) 16232-16237. doi:10.1016/j.ceramint.2017.08.205

[39] O. Porat, I. Riess, Defect chemistry of $Cu_{2-y}O$ at elevated temperatures. Part II: Electrical conductivity, thermoelectric power and charged point defects, Solid State Ionics, 81 (1995). doi:10.1016/0167-2738(95)00169-7

[40] Y. Wang, P. Miska, D. Pilloud, D. Horwat, F. Mücklich, J.F. Pierson, Transmittance enhancement and optical band gap widening of $Cu_2O$ thin films after air annealing, Journal of Applied Physics, 115 (2014) 073505. doi:10.1063/1.4865957

[41] Syed Farid Uddin Farhad, David Cherns, Structural, Optical and Electrical properties of nanocrystalline $Cu_2O$ thin films grown by Pulsed Laser Deposition, Paper# 9AM-J8.30 Abstract book document# 1868681 in: 2014 MRS Spring Meeting and Exhibition, Material Research Society(MRS), San Francisco, California, USA. http://dx.doi.org/10.17632/whksg7scsx.1

[42] L. Debbichi, M.C. Marco de Lucas, J.F. Pierson, P. Krüger, Vibrational Properties of CuO and $Cu_4O_3$ from First-Principles Calculations, and Raman and Infrared Spectroscopy, The Journal of Physical Chemistry C, 116 (2012) 10232-10237. doi:10.1021/jp303096m

[43] S.F.U. Farhad, N.I. Tanvir, M.S. Bashar, M.S. Hossain, M. Sultana, N. Khatun, Facile synthesis of oriented zinc oxide seed layer for the hydrothermal growth of zinc oxide nanorods, Bangladesh Journal of Scientific and Industrial Research, 53 (2018) 233. doi:10.3329/bjsir.v53i4.39186

[44] S.B. Ogale, P.G. Bilurkar, N. Mate, S.M. Kanetkar, N. Parikh, B. Patnaik, Deposition of copper oxide thin films on different substrates by pulsed excimer laser ablation, Journal of Applied Physics, 72 (1992) 3765-3769. doi:10.1063/1.352271

[45] A. Subramaniyan, J.D. Perkins, R.P. O'Hayre, S. Lany, V. Stevanovic, D.S. Ginley, A. Zakutayev, Non-equilibrium deposition of phase pure $Cu_2O$ thin films at reduced growth temperature, APL Materials, 2 (2014) 022105. doi:10.1063/1.4865457

[46] B. Bera, A. Chakraborty, T. Kar, P. Leuaa, M. Neergat, Density of States, Carrier Concentration, and Flat Band Potential Derived from Electrochemical Impedance Measurements of N-Doped Carbon and Their Influence on Electrocatalysis of Oxygen Reduction Reaction, The Journal of Physical Chemistry C, 121 (2017) 20850-20856. doi:10.1021/acs.jpcc.7b06735

[47] R. Garuthara, W. Siripala, Photoluminescence characterization of polycrystalline n-type $Cu_2O$ films, Journal of Luminescence, 121 (2006) 173-178. doi:10.1016/j.jlumin.2005.11.010

[48] L. Zhang, L. McMillon, J. McNatt, Gas-dependent bandgap and electrical conductivity of $Cu_2O$ thin films, Solar Energy Materials and Solar Cells, 108 (2013) 230-234. doi:10.1016/j.solmat.2012.05.010

[49] X. Chen, D. Parker, M.-H. Du, D.J. Singh, Potential thermoelectric performance of hole-doped $Cu_2O$, New Journal of Physics, 15 (2013) 043029. doi:10.1088/1367-2630/15/4/043029

# Supplementary Material

# Pulsed laser deposition of single phase n- and p-type Cu$_2$O thin films with low resistivity

*Syed Farid Uddin Farhad[1,2,3,4*], David Cherns[1], James Smith[2], Neil Fox[1,2], and David Fermín[3]*

[1]H.H. Wills Physics Laboratory, School of Physics, University of Bristol, BS8 1TL, UK

[2]Diamond Laboratory, School of Chemistry, University of Bristol, BS8 1TS, UK

[3]Electrochemistry Laboratory, School of Chemistry, University of Bristol, BS8 1TS, UK

[4]Industrial Physics Division, BCSIR Labs, Dhaka, Bangladesh Council of Scientific and Industrial Research (BCSIR), Dhaka 1205, Bangladesh

[*]Corresponding Author: sf1878@my.bristol.ac.uk ; s.f.u.farhad@bcsir.gov.bd

**Copper oxide films grown with O$_{2pp}$ = 10 and 3 mTorr at T$_{sub}$ = 25 - 400 $^0$C :**

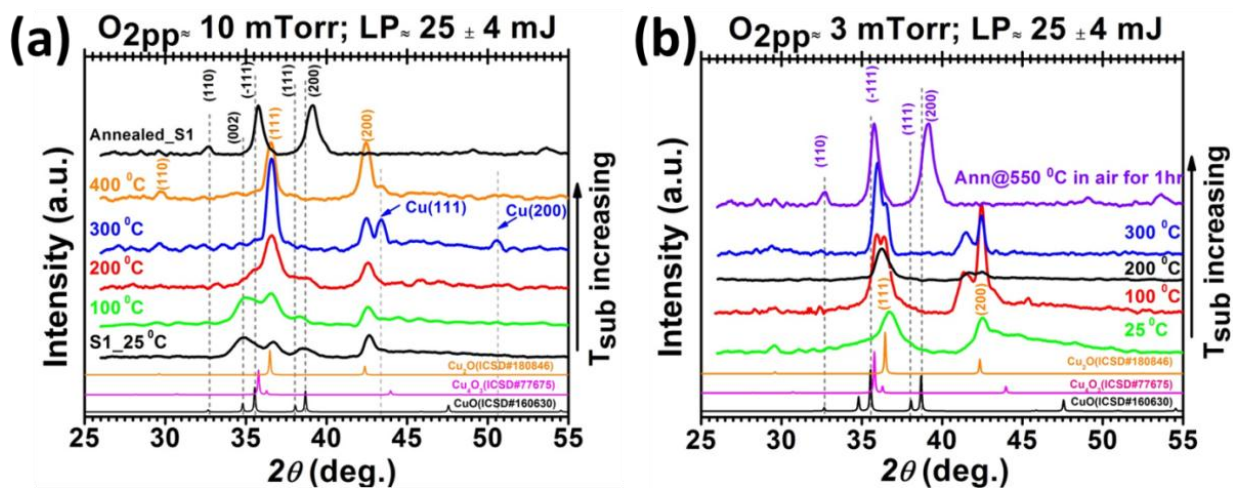

**Fig. S1.** (color online) XRD patterns (vertically offset for clarity) of copper oxide thin films deposited onto quartz substrate with a constant laser pulse energy (LP ≈ 25±4 mJ) using (a) O$_{2pp}$ = 10 mTorr, and (b) O$_{2pp}$ = 3 mTorr as a function of substrate temperature. The dotted and solid vertical lines indicate the Bragg reflection peaks of CuO and Cu respectively.

**$O_{2pp}$ = 10 mTorr**:

Thin films deposited at $T_{sub} \approx 100\ ^0C$, show a mixture of CuO and $Cu_2O$ phases. In contrast, films grown at $T_{sub} \geq 200\ ^0C$ exhibit strong $Cu_2O(111)$ and $Cu_2O(200)$ reflection peaks but no clear evidence of CuO and $Cu_4O_3$ phases (see the respective ICSD patterns given in the figure). For films grown at $T_{sub} \geq 300\ ^0C$, there is also evidence of metallic Cu with (111) and (200) planes in XRD patterns. In addition, $Cu_2O$ phase is seen to be increasing with the expense of CuO phase, suggesting that an elevated temperature might be more favourable for PLD ablated adatoms to form single phase $Cu_2O$ but it also suggests that even $O_{2pp}$ = 10 mTorr is not enough to suppress Cu within $300\ ^0C \leq T_{sub} \leq 400\ ^0C$ regime (see blue and orange curve in Fig. S1a). The likely reason of the absence of CuO but the presence of $Cu_2O$ and Cu most probably due to the dissolving of O species from the Cu-O films in $300\ ^0C \leq T_{sub} \leq 400\ ^0C$ regime. These results suggest that scope of controlling the phase purity is less certain in case the of oxygen rich PLD ambient.

**$O_{2pp}$ = 3 mTorr**:

Thin films deposited at $T_{sub} \approx 25\ ^0C$, exhibit strong $Cu_2O$ only with (111) and (200) but those grown at $100\ ^0C \leq T_{sub} \leq 300\ ^0C$ contains $Cu_2O$ phase along with $Cu_xO_y$ [1].

**$Cu_2O$ films grown at $T_{sub}$ = 25 $^0C$ on amorphous, polycrystalline and crystalline substrate:**

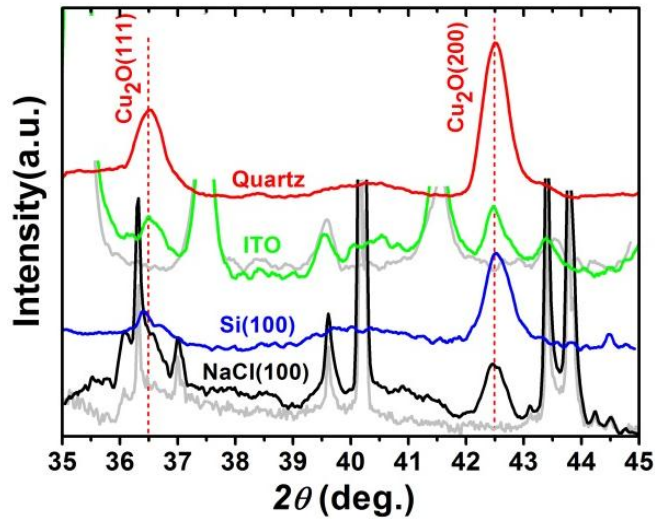

**Fig. S1.1.** (color online) XRD patterns (vertically offset for clarity) of $Cu_2O$ films deposited on various substrates at $O_{2pp}$ = 5 mTorr_$T_{sub} \approx$ 25 °C. XRD from blank ITO and NaCl(100) substrates are shown by fade curves and included in the graph for comparison purposes. Reference peak positions of pure $Cu_2O$ are indicated by vertical dotted lines.

Clearly, XRD patterns of thin films deposited on quartz glass, polycrystalline ITO, (100) oriented Si and (100) oriented NaCl substrate revealed that all films are polycrystalline in nature irrespective of the type of substrate used. These diffractogram exhibit 111 and 200 reflections of $Cu_2O$ phase with an average lattice parameter, $a$~4.26±0.02 Å (cf. $a$~4.27 Å for bulk $Cu_2O$), no evidence of other Cu, $Cu_4O_3$ and CuO phases. Phase pure $Cu_2O$ films grown at $T_{sub}$≈ 25 °C on (100) oriented substrates exhibit 200 texturing as evident from Fig. S1.1.

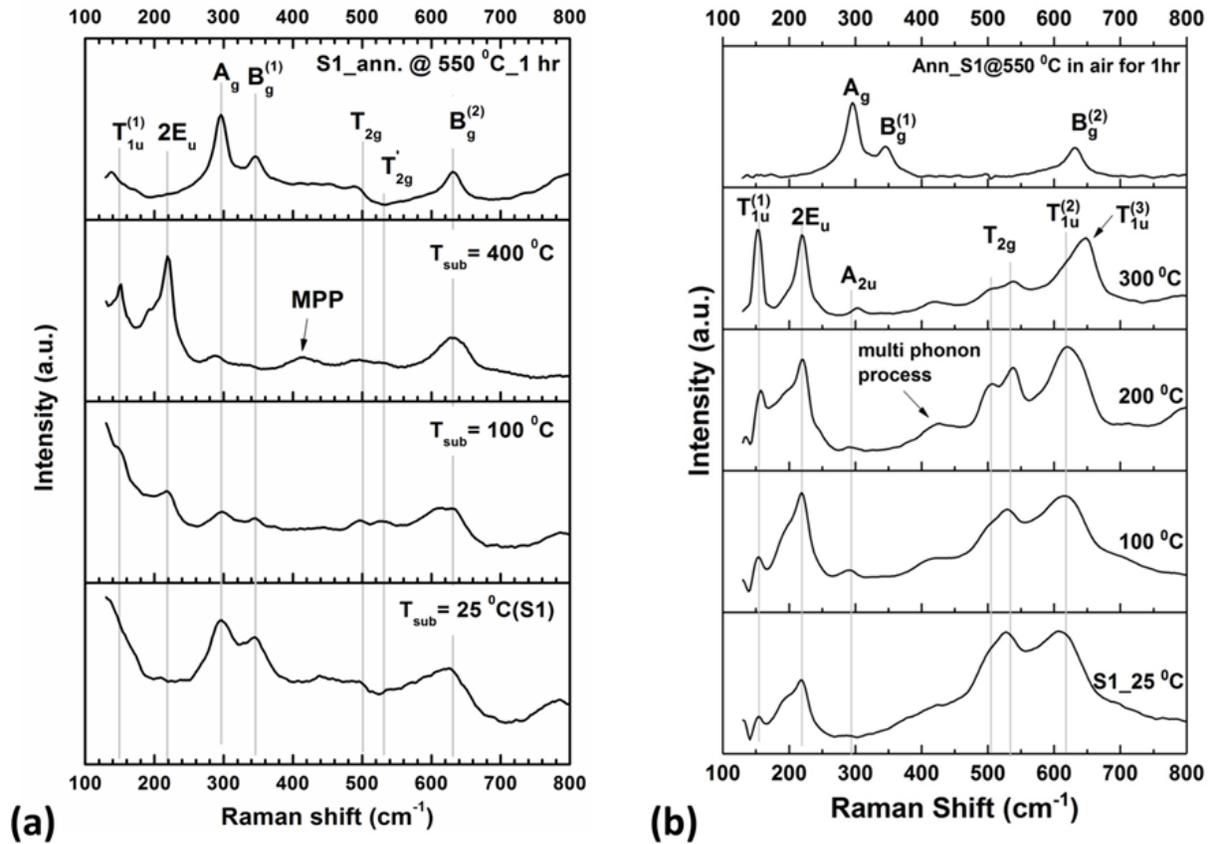

**Fig. S2.** Raman spectra of copper oxide thin films deposited onto quartz substrate with a constant laser pulse energy (LP ≈ 25±4 mJ) using (a) $O_{2pp}$ = 10mTorr, and $O_{2pp}$ = 3mTorr (b) as a function of substrate temperature. The vertical lines indicate the reference vibrational modes of copper oxide [1,2].

In order to understand the crystallographic nature of PLD grown $Cu_2O$ films, texture coefficient (Fig. S3a), and crystallite size (Fig. S3b) were analysed as a function of $O_{2pp}$ in both RT- and HT-grown films. Texturing coefficient (f) is calculated from the following formula [2] considering (111) and (200) are the only planes lying parallel to the substrate and hence diffracting strongly:

$$f = 1 - 2y/(x+y) \quad\quad\quad (S1)$$

where, $x = \frac{I(111)}{I(200)}$ , $y = \frac{I_0(111)}{I_0(200)}$ ; I(hkl) and $I_0$(hkl) are the intensities of (hkl) X-ray reflection planes of as grown thin films and bulk $Cu_2O$ respectively. (XRD of bulk $Cu_2O$ powder scrapped off from PLD target is shown in Fig. S7.1 below). When f→-1, thin films are highly (200) textured and crystallites are dominantly (111) orientated when f→1. One can see that the RT- and HT-grown films approached towards highly {100} and {111} texturing, respectively, with increasing $O_{2pp}$ up to 7 mTorr.

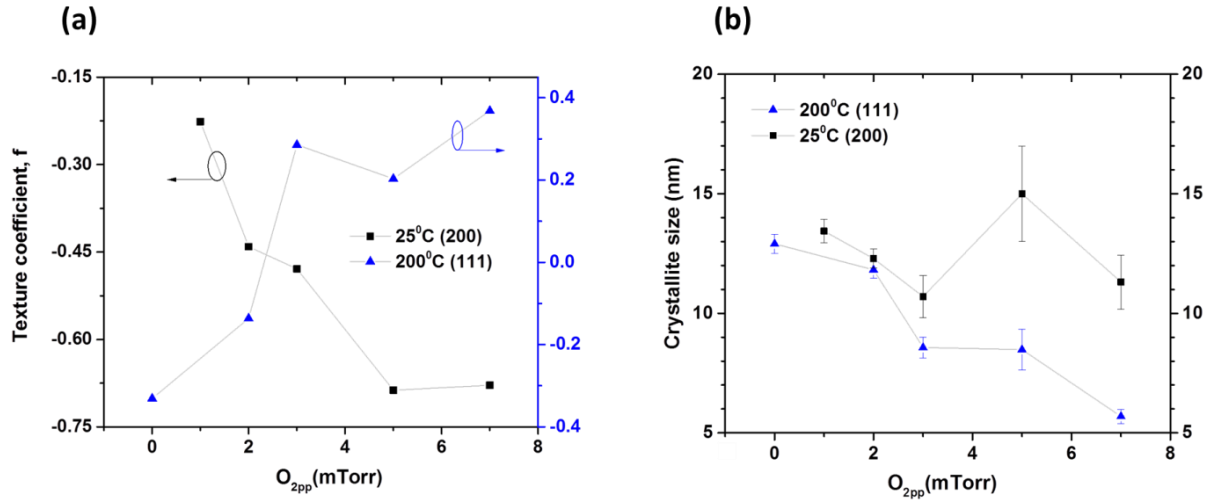

**Fig. S3**. Texture coefficient (a), and average crystallite domain size(b) of $Cu_2O$ films deposited at $T_{sub} \approx 25\ ^0C$ (denoted by ■) and $T_{sub} \approx 200\ ^0C$ (denoted by ▲) as a function of $O_{2pp}$.

Fig. S3b shows the variation of the average crystallites domain size as a function of $O_{2pp}$. Both for RT- and HT-grown films, grain growth is negatively affected with increasing $O_{2pp}$ ; for example, in case of HT-grown films, crystallite size start to decrease monotonically from ~12 nm with no $O_{2pp}$ down to ~5 nm with $O_{2pp} \approx 7$ mTorr; suggesting that $O_{2pp}$ must have greater impact on controlling growth rate as well as films microstructure. The likely reason is that higher level of $O_{2pp}$ present during PLD scatter, attenuate and thermalize the ablation plume; consequently lowering the kinetic energy of particle arriving the substrate surface. Therefore, the lower the kinetic energy of adatoms, the slower the grain growth as well as the deposition rate. Furthermore, the crystallite size in HT-grown films are always found to be smaller than those of the RT-grown films irrespective of the $O_{2pp}$ level, indicating that the growth rate in RT-films is faster than HT-films, which corroborates the results presented Fig. 6 of the main text.

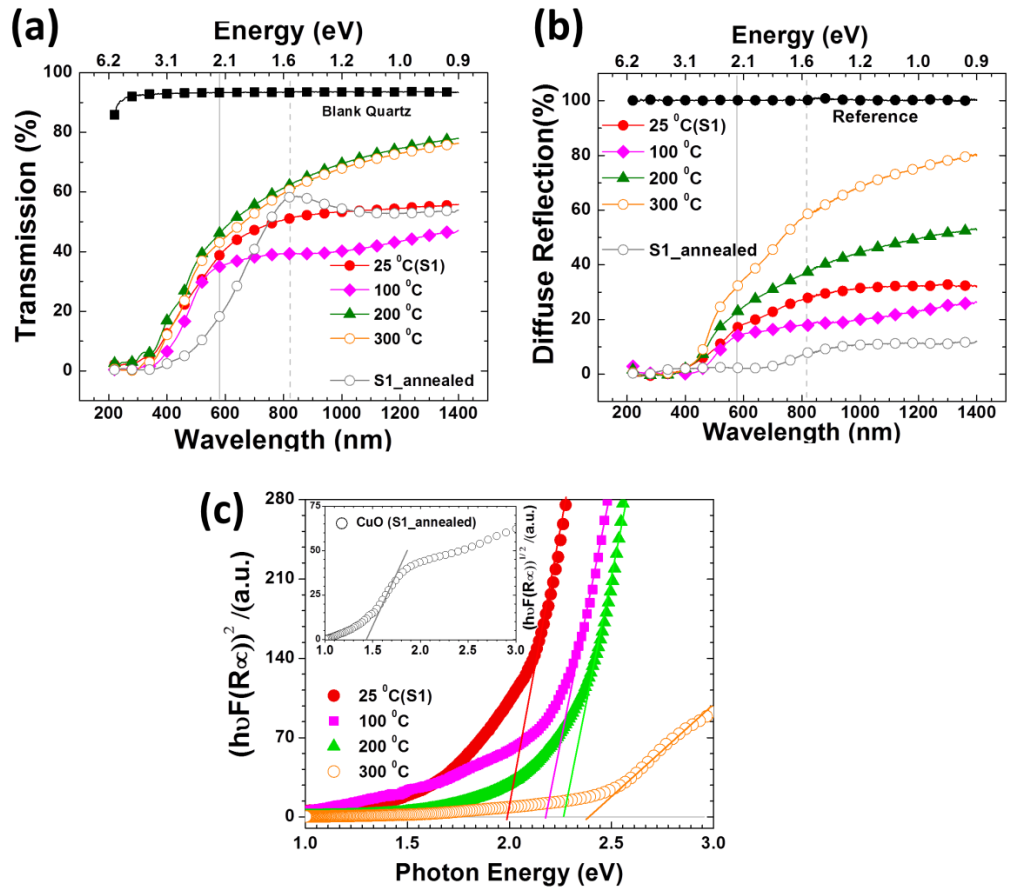

**Fig. S4.** (color online) Transmission (a), diffuse reflection (b) and Tauc plot (c) of PLD grown $Cu_2O$ films as a function of substrate temperature. Tauc plot of the CuO (S1_annealed) film is shown in the inset.

PLD films grown at $O_{2pp}$ = 3 mTorr_$T_{sub}$ = 300 $^0$C contain predominant $Cu_xO_y$ phase (see Fig. S1b) and exhibited a larger bandgap ($E_g$) ~ 2.45 eV (direct) compared to single phase $Cu_2O$ film ($E_g$ =1.96-2.20 eV) and RT-grown sample (S1) annealed in air at 550 $^0$C for 1 hour exhibited a bandgap ~1.45 eV (indirect) due to complete conversion to CuO.

The electrical resistivity measured by four point collinear probe exhibited gradual increase in resistivity with increasing substrate temperature at a fixed $O_{2pp}$. Fig. S5 (below) shows the variation of resistivity (denoted by solid circle (●)) and thickness (denoted by solid rectangle (■)) as a function of $T_{sub}$ at constant $O_{2pp} = 3$ mTorr. At fixed $O_{2pp} = 3$ mTorr, the RT(25 $^0$C) -grown film exhibited ~ 6 orders of magnitude lower resistivity (23 ± 1 mΩ.cm) than those of the films grown at 100 $^0$C ≤ $T_{sub}$ ≤ 300 $^0$C (their resitivities are found in the range 5 – 30 kΩ.cm).

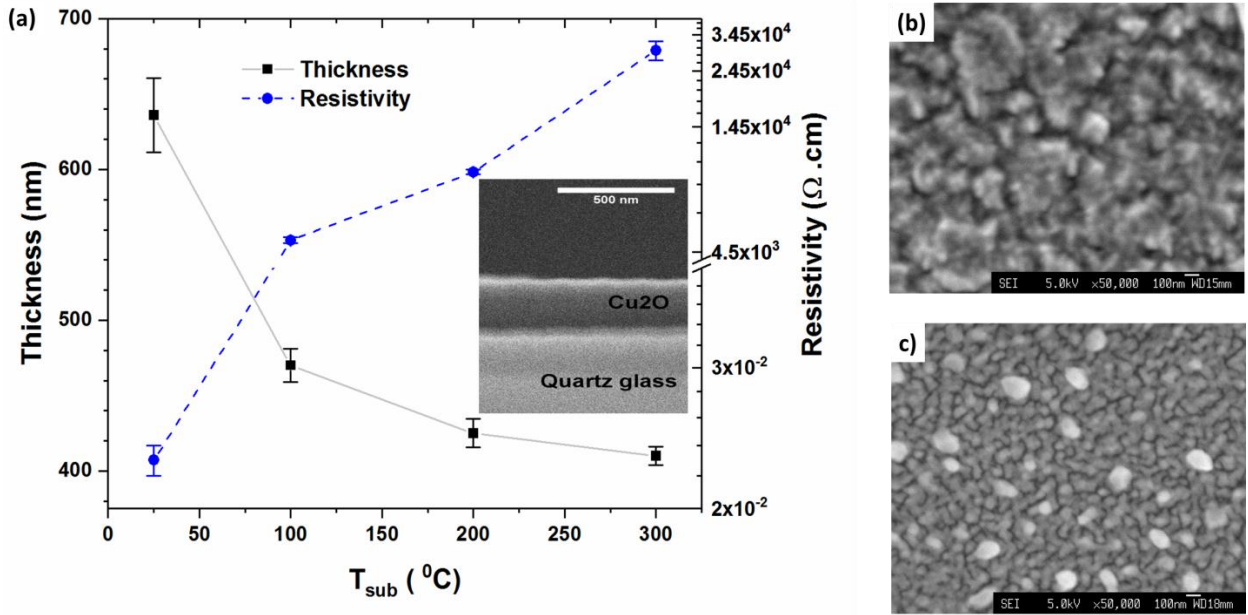

**Fig. S5.** Variation in electrical resistivity and thickness of $Cu_2O$ thin film grown at $O_{2pp} = 3$ mTorr as a function of substrate temperature (a); Plane SEM view of $Cu_2O$ film deposited at $T_{sub}$ ≈ 25 $^0$C (b) and at $T_{sub}$ ≈ 300 $^0$C (c). The thickness data presented in (a) are the average value of VASE and SEM measurements (inset shows a typical cross-sectional SEM view of a sample grown at $T_{sub}$ ≈ 300 $^0$C).

Notice the film thickness is consistently decreasing from ~ 635 ± 25 nm to ~ 410 ± 6 nm when substrate temperature is elevated from $T_{sub} = 25$ $^0$C(RT) to $T_{sub} = 300$ $^0$C. The estimated film thickness of $Cu_2O$ with 2 mTorr ≤$O_{2pp}$≤ 7 mTorr was found to be roughly in the range 550 – 625 nm and 500 – 600 nm for 25 $^0$C - and 200 $^0$C -grown films respectively (data not shown here)). Intriguingly, grains in RT-grown film are significantly larger than those of $T_{sub}$ ≈ 300 $^0$C- grown films as evident from SEM micrographs (cf. Fig. S5b and S5c), suggesting that the columnar like grain (with a large density of voids) growth in the former and lateral growth in the latter might be playing a vital role in decreasing growth rate with increasing growth temperature. The higher resistivity of thin films grown at elevated $T_{sub}$ suggesting that samples being investigated are in general more oxygen rich as well as stoichiometric compared to RT(25 $^0$C)- grown thin films. This observation corroborates the analyses results of XRD and PL (see Fig. 2 and Fig. 4 in the main text).

**Surface morphology of PLD grown Copper oxide thin films on different substrates:**

The FE-SEM investigated surface morphologies of copper-oxide thin films grown on quartz glass, ITO, NaCl(100), and AZO substrates at various deposition and processing conditions are presented below:

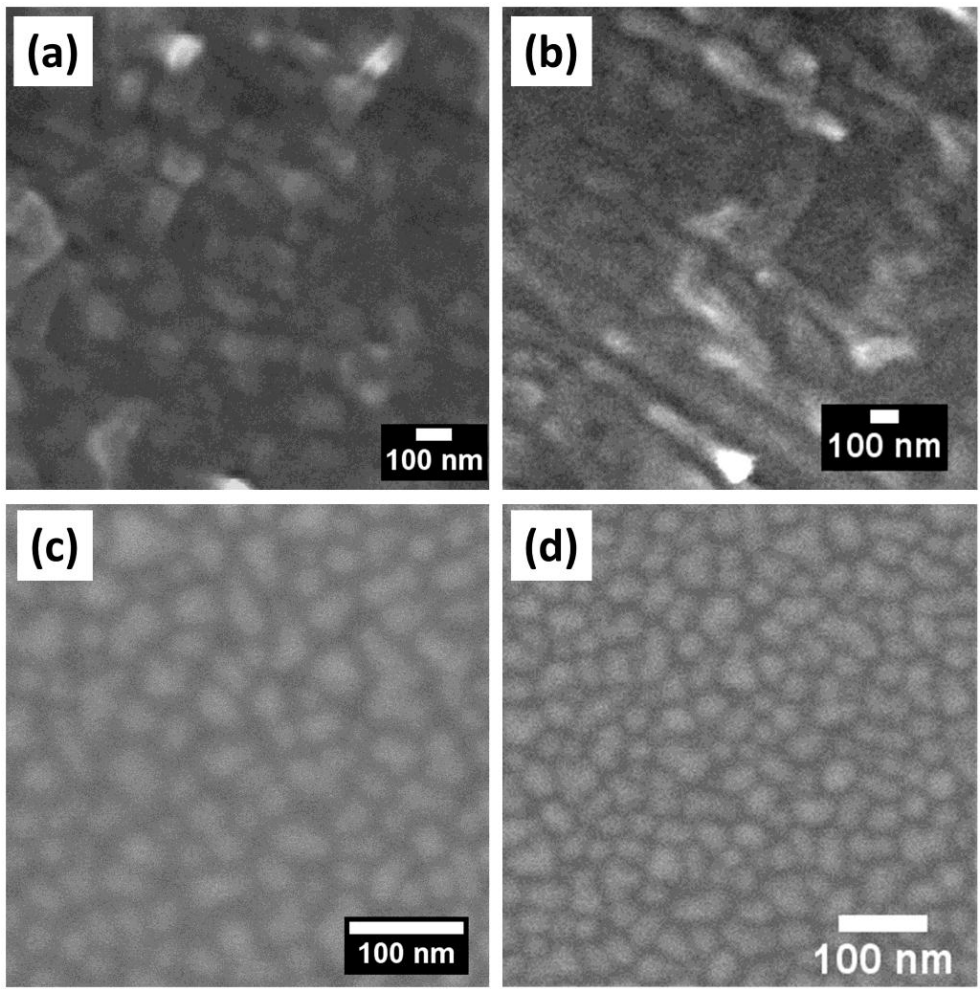

**Fig. S5.1.** SEM micrographs of copper oxide thin films deposited at (a) $O_{2pp}$ = 1 mTorr_$T_{sub}$≈ 25 ºC ,(b) $O_{2pp}$ = 3 mTorr_$T_{sub}$≈ 25 ºC and (c) $O_{2pp}$ << 1 mTorr_$T_{sub}$≈ 200 ºC and (d) $O_{2pp}$ << 1 mTorr_$T_{sub}$≈ 300 ºC on quartz glass.

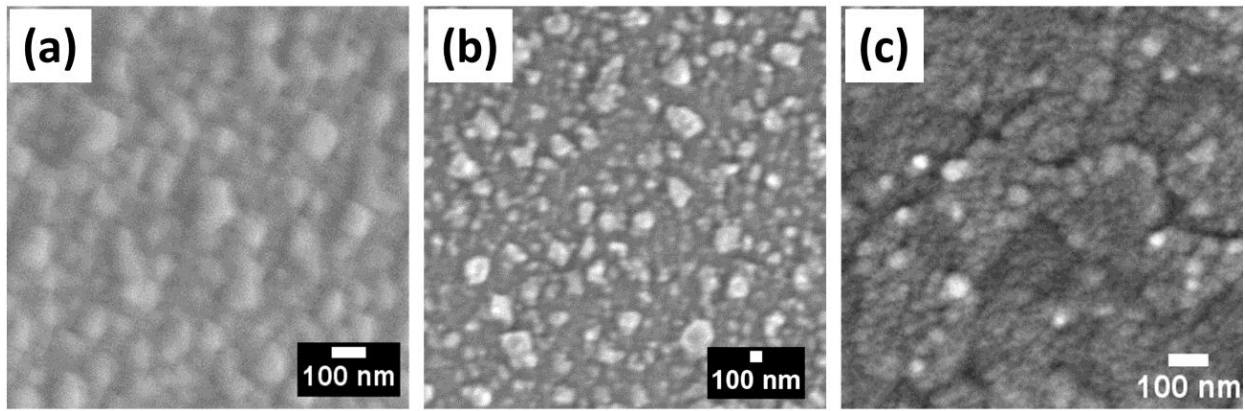

**Fig. S5.2.** SEM micrographs of Cu$_2$O thin films deposited at T$_{sub}$≈ 25 °C with (a) 2 mTorr O$_{2pp}$ and (b) 5 mTorr O$_{2pp}$ onto commercial ITO coated soda lime glass (c).

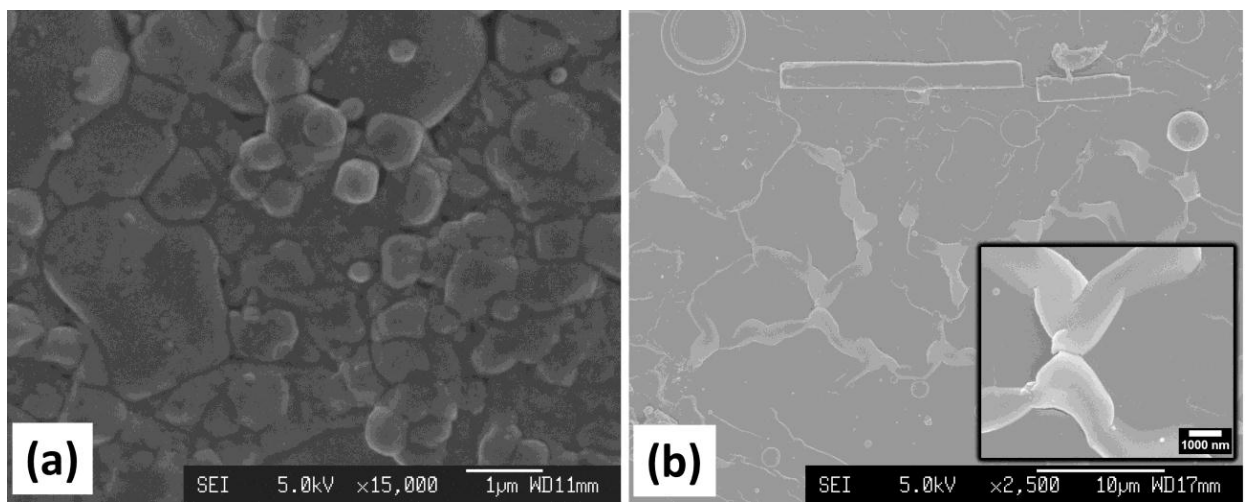

**Fig. S5.3**: SEM micrographs of Cu$_2$O thin films (a) deposited with O$_{2pp}$= 5 mTorr , LP ≈ 30±2 mJ  at T$_{sub}$≈ 200 °C on NaCl(100) substrate (b). A magnified image of the blank NaCl(100) are shown in the right figure with scale bar 1 μm.

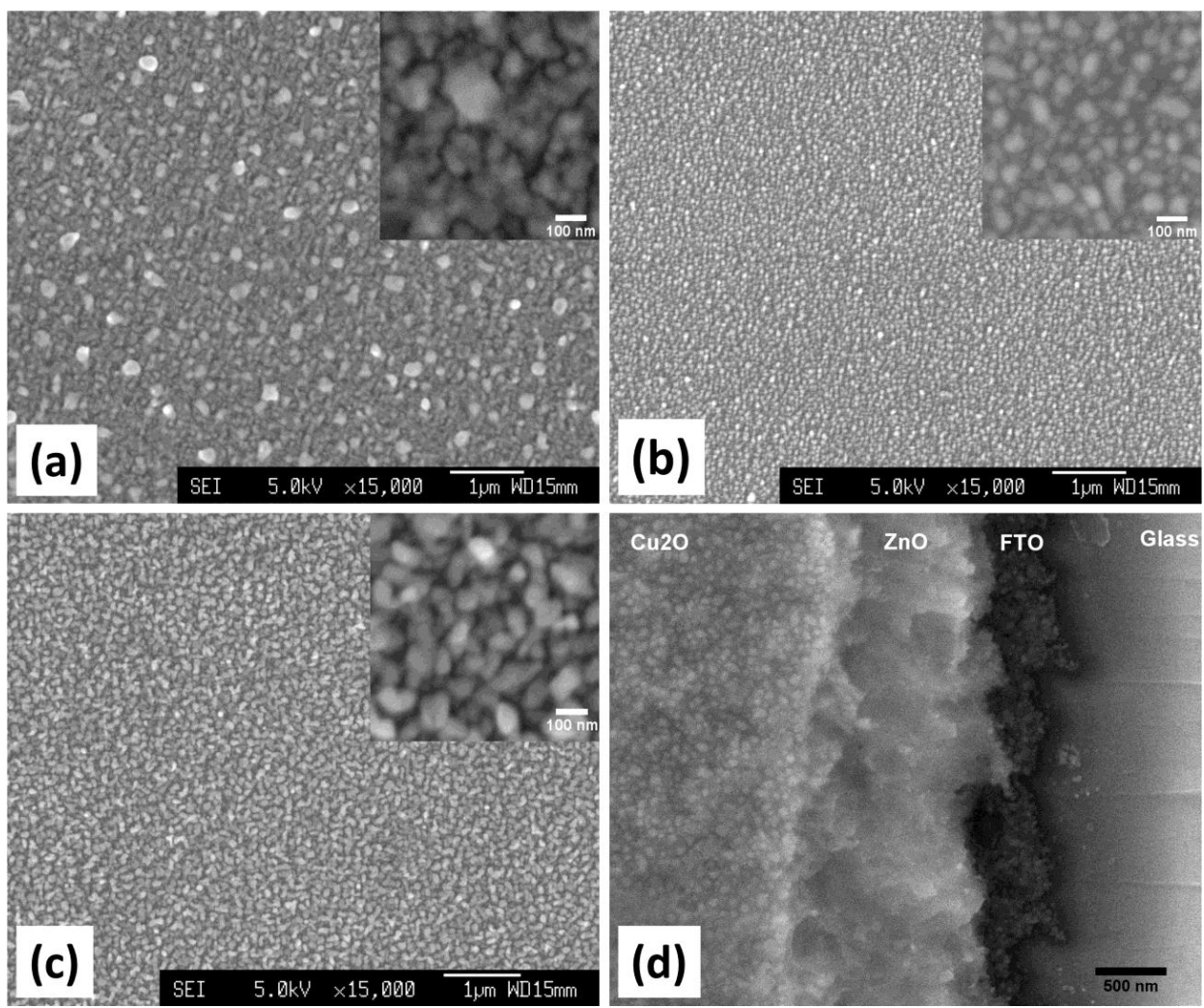

**Fig. S5.4.** SEM (Plane view) micrographs of $Cu_xO_y$ thin films grown at $T_{sub} \approx 300\ ^oC$ with (a) 3 mTorr $O_{2pp}$, (b) 5 mTorr $O_{2pp}$, and (c) 10 mTorr $O_{2pp}$, respectively onto Al-doped ZnO coated (PLD grown at 10 mTorr_$T_{sub} \approx 400\ ^0C$) soda lime glass. Deposition time ~ 45 min for samples (a) – (c). (d) SEM (cross-sectional view, ~$7^0$ titled ) micrograph of $Cu_xO_y$ thin films grown at 3 mTorr $O_{2pp}$_$T_{sub} \approx 300\ ^oC$ (Deposition time ~ 45 min) onto ZnO coated (PLD grown at 10 mTorr_$T_{sub} \approx 300\ ^0C$, Deposition time 30 min) FTO glass.

**XRD of $Cu_2O$ ceramic PLD target (purity ~ 99.95%) and ICSD powder diffraction files of $Cu_2O$, $Cu_4O_3$ and CuO phases:**

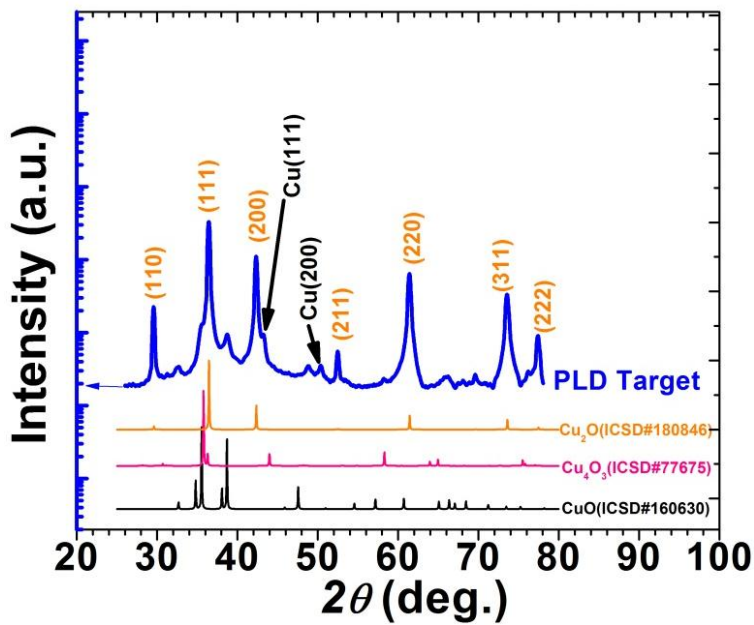

**Fig. S7.1.** (color online) XRD patterns of PLD target used for deposition of copper oxide thin films (log scaled indicating by arrow in the left) together with standard ICSD patterns of three different copper-oxide compounds (linear vertical scale in the right).

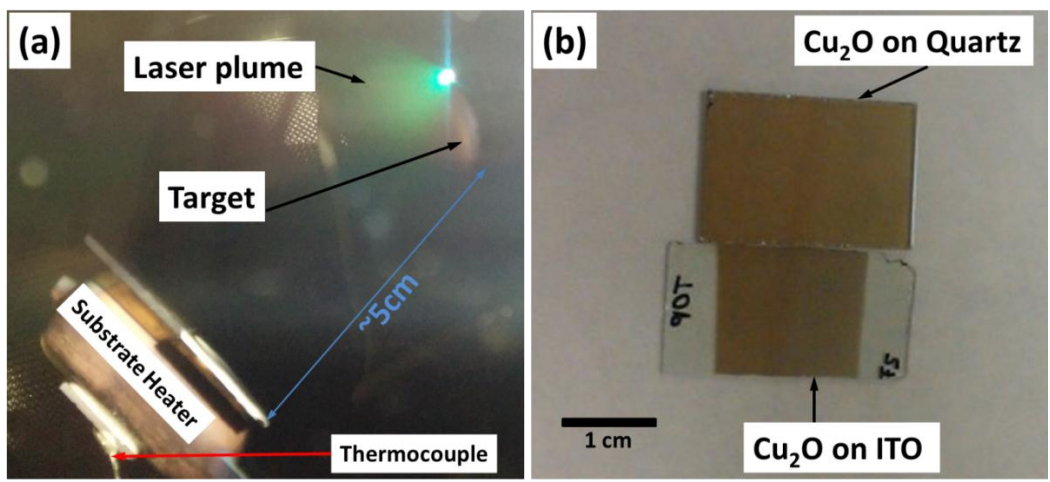

**Fig. S7.** (a) Photograph taken during the synthesis of $Cu_2O$ film on two different substrates in a single deposition session. (b) Photograph of $Cu_2O$ thin films on ITO and Quartz substrate after deposition. Part of ITO was masked with PTFE tape during deposition for selective ohmic contact required for subsequent electrochemical impedance measurements.

**Hall mobility measurements**:

Measurements of the Hall voltage and sheet resistance of samples were carried out using the van der Pauw configuration as per the standard procedure set out by NIST Physical Measurement Laboratory [3]. Hall voltage measurements were performed using a 1 Tesla (T) permanent magnet, with the polarity of the magnetic field (B(+) and B(-)) being reversed manually through sample stage orientation relative to the magnetic field. All measurements were taken by sourcing a range of currents and measuring the voltage change (see inset in Fig. S8 (left)). The resultant voltage step functions allow the offset voltage to be measured and taken into account when calculating the Hall voltage and mobilities. The offset voltage is typically a product of non-ideal Ohmic contact, asymmetric contact placement, or irregular sample shape. Joule heating of the sample during testing was also reduced by using the minimum permitted source current thereby minimising the thermal contribution in the measurement, following the ASTM recommendation ref: F46. All electrical measurements were performed at room temperature within a light-tight Faraday cage.

Hall voltages, $V_H$, were measured with varying source current, I. The four measurement configurations were averaged to form ($V_H$/I) which was used for calculating mobility via the equation, $\mu_H = (V_H/I) \times 10^4 \times B^{-1} \times (R_s)^{-1}$, where magnetic field B is in T and $R_s$ is the sheet resistance in ohm/square. The large offset voltage (Fig. S10) seen in the measurements were subtracted to reveal the actual $V_H$. $R_s$ values measured in 4-point probe were found to be of the order of ± 2-5 % of that of the van der Pauw measurements, this is thought to be primarily due to the different nature of the electrical contact arrangement and the sample size.

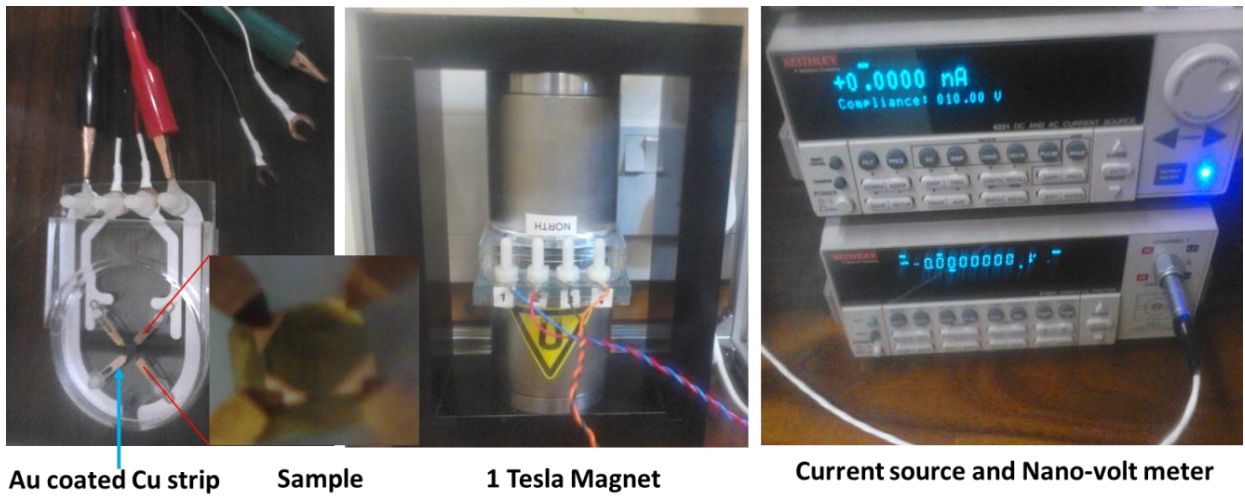

**Fig. S8.** (color online) A precision home-built Hall mobility measurement setup used in this study.

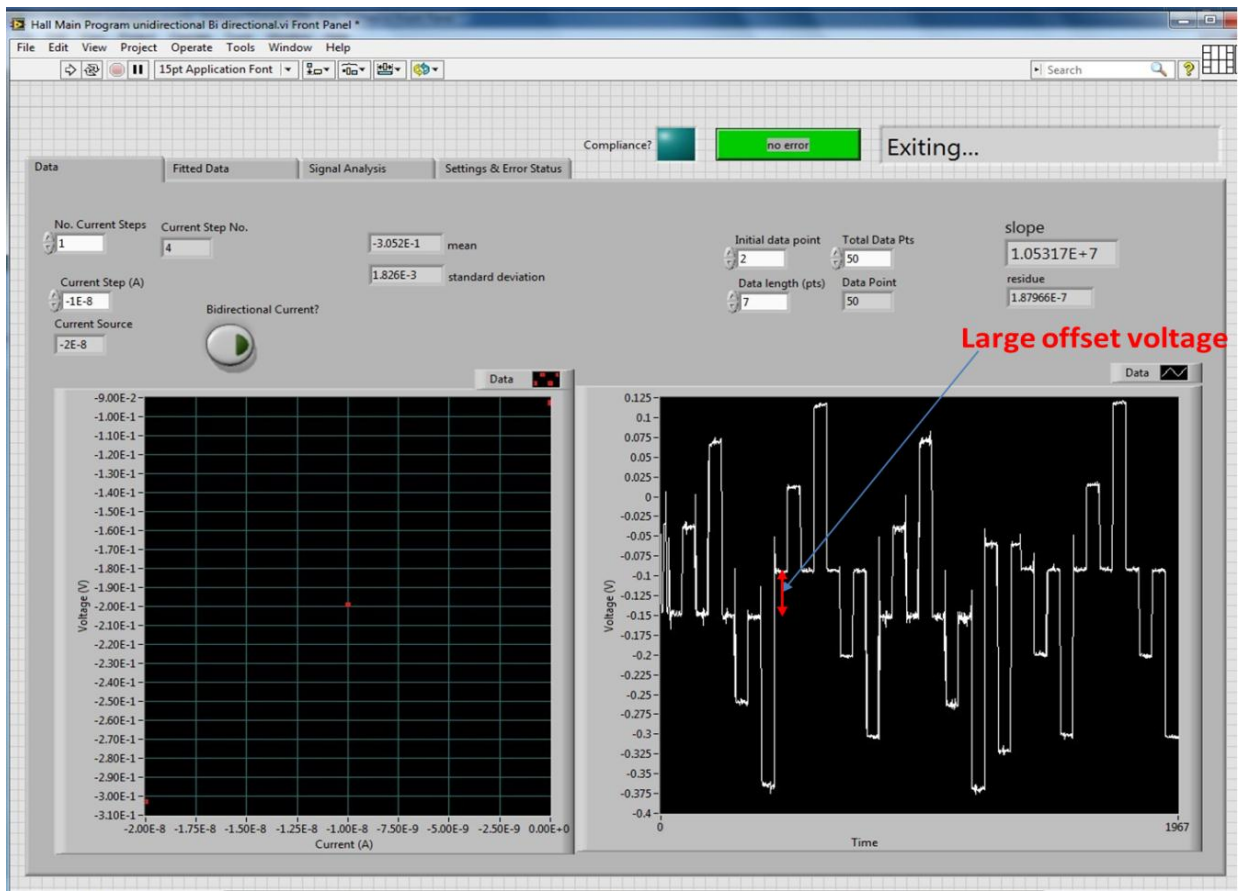

**Fig. S9.** (color online) An example of a large offset voltage (LabVIEW program display) in a typical Hall measurement setup due to non-ideal contact and sample shape.

**Electrochemical Impedance Spectroscopy (EIS):**

To evaluate electrochemical impedance parameter the grequency response analysis (FRA) was perfomed by appling a set of constant bias voltage over a narrow potential window (-0.25 to 0.05 V *vs* RHE) as mentioned in the experimental section 2, and a standby potential ca. 0.01 V (*vs* RHE) was maintained after each measurement to avoid possible formation of native CuO atop the film. This was confirmed by doing CV before and after FRA as shown Fig. S10 evident from the low capacitive current (few µA) within the potential scanning range. It is worth mentioning here that the experimental electrochemical impedance parameters are fitted to a suitable RQ (Q is the constant phase element (CPE) having relation with $C = (Q*R)^{(1/n)}/R$ and $Q \approx C$ as $n \geq 0.9$) equivalent circuit combination in high frequency region as in Fig. S11(below) to evaluate true capacitances. This capacitance is due to space charge region at the copper oxide|electrolyte interface and capacitances at different bias voltage are used to construct the Mott-Schottky plot shown in Fig. 7a in the main text. The Mott-Schottky realtion was used to work out $V_{fb}$ and intrinsic carrier density ($N_x$, where x = *d* for n-type and *a* for p-type). The values determined in this way, for desired RT-grown samples, are included in the Table 1( see main text) along with other properteis as a function of $O_{2pp}$. The flatband potential converted to vacuum scale using similar method describe in the litterature [4] and also depicted in Fig. 7 (main text) to indentify conduction band (CB) and valence band (VB) with a common reference.

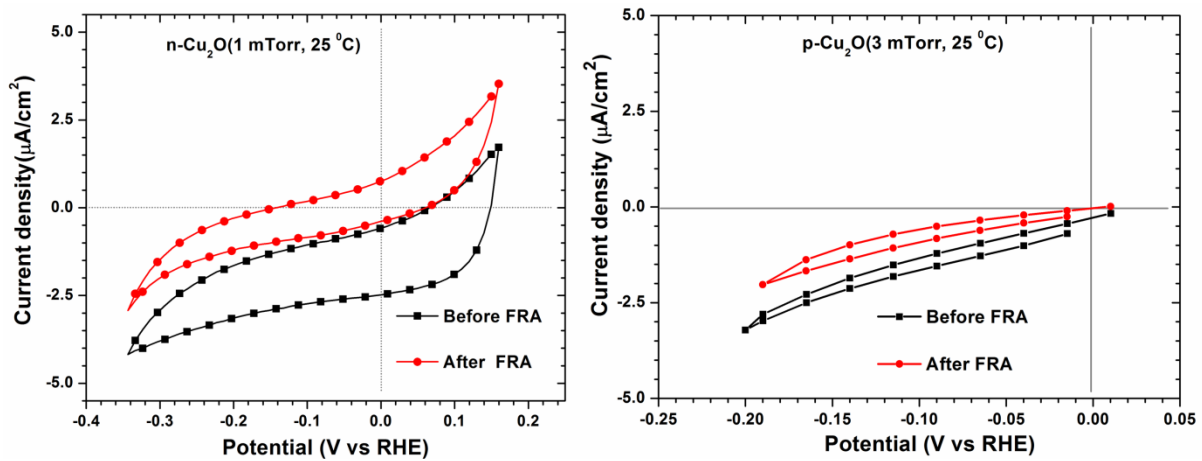

**Fig. S10.** Cyclic voltammograms of n- $Cu_2O$/ITO (left) and p-type $Cu_2O$/ITO (right) electrodes taken before and after frequency response analyses confirm that there was no-faradic process occurred during EIS measurement.

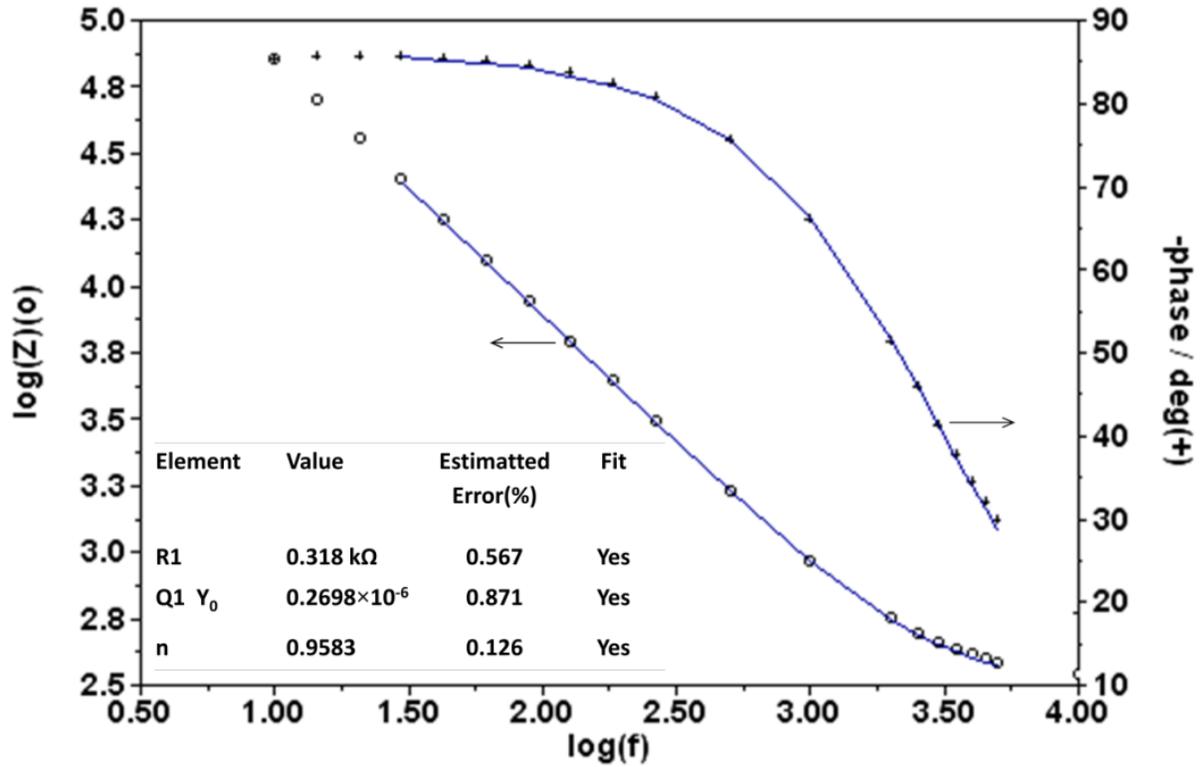

**Fig. S11.** Bode plot of EIS measurements of a RT-grown $Cu_2O$ film on ITO in argon saturated 0.1M CH3COONa electrolyte at -0.1 V *vs* RHE, the solid line represent the fits using an RQ equivalent(Q≈ C as n≥0.9). First few data points are eliminated to minimize the error in the best fitted impedance parameters (see inset table).

**Calculation of effective density of states ( $D(E_f)$ ) from EIS data:**

The total double-layer capacitance ($C_T$) of copper oxide|electrolyte system can be calculated from the imaginary impedance ($Z_{imag}$) using equation [5]:

$$C_T = -1/(2\pi f Z_{imag}) \qquad (S2)$$

where, f is the frequency. The space charge capacitance ($C_{sc}$) and Helmohltz capacitance ($C_H$)[6] can be related to $C_T$ by following equation:

$$1/ C_T = 1/ C_{sc} + 1/ C_H \qquad (S3)$$

For, $Cu_2O$/ITO electrode grown at 25 $^0C$ with $O_{2pp}$ = 3 mTorr, using equation (S2) we got, $C_T$ = 11. 25 µF/cm$^2$ (where, f = 29.76 Hz  and - $Z_{imag}$ ≈ 6790 Ohm) at -0.21 V *vs* Ag/AgCl(≈0 V *vs* RHE) considering similar approach described in ref. [5].

Assuming $C_H$ >> $C_{sc}$ ; $C_{sc}$ ≈ $C_T$ ≈ 11. 25×10$^{-2}$ F/m$^2$

Therefore, the effective density states, $D(E_f) = (C_{sc})^2 / [\varepsilon_0.\varepsilon.e^2] \approx 1.35 \times 10^{21}$ cm$^{-3}$/(eV).

[Or, incase of highly doped electrodes, assuming $C_H$ = 20 µF/cm$^2$ [5] equation (S3) yields, $C_{sc}$ ≈ 25.70 µF/cm$^2$; $D(E_f) \approx 7.06 \times 10^{21}$ cm$^{-3}$/(eV)] where, $\varepsilon_0$ is permittivity of free space (8.854×10$^{-12}$ F.m$^{-1}$), $\varepsilon$ is the relative dielectric constant(~6.6 for $Cu_2O$ [7]) and *e* is the electronic charge ~1.60×10$^{19}$ C.

Similarly, we estimated effective density of sates for other samples.

**Fabrication of solid p-n junctions using p- and n-type Cu$_2$O:**

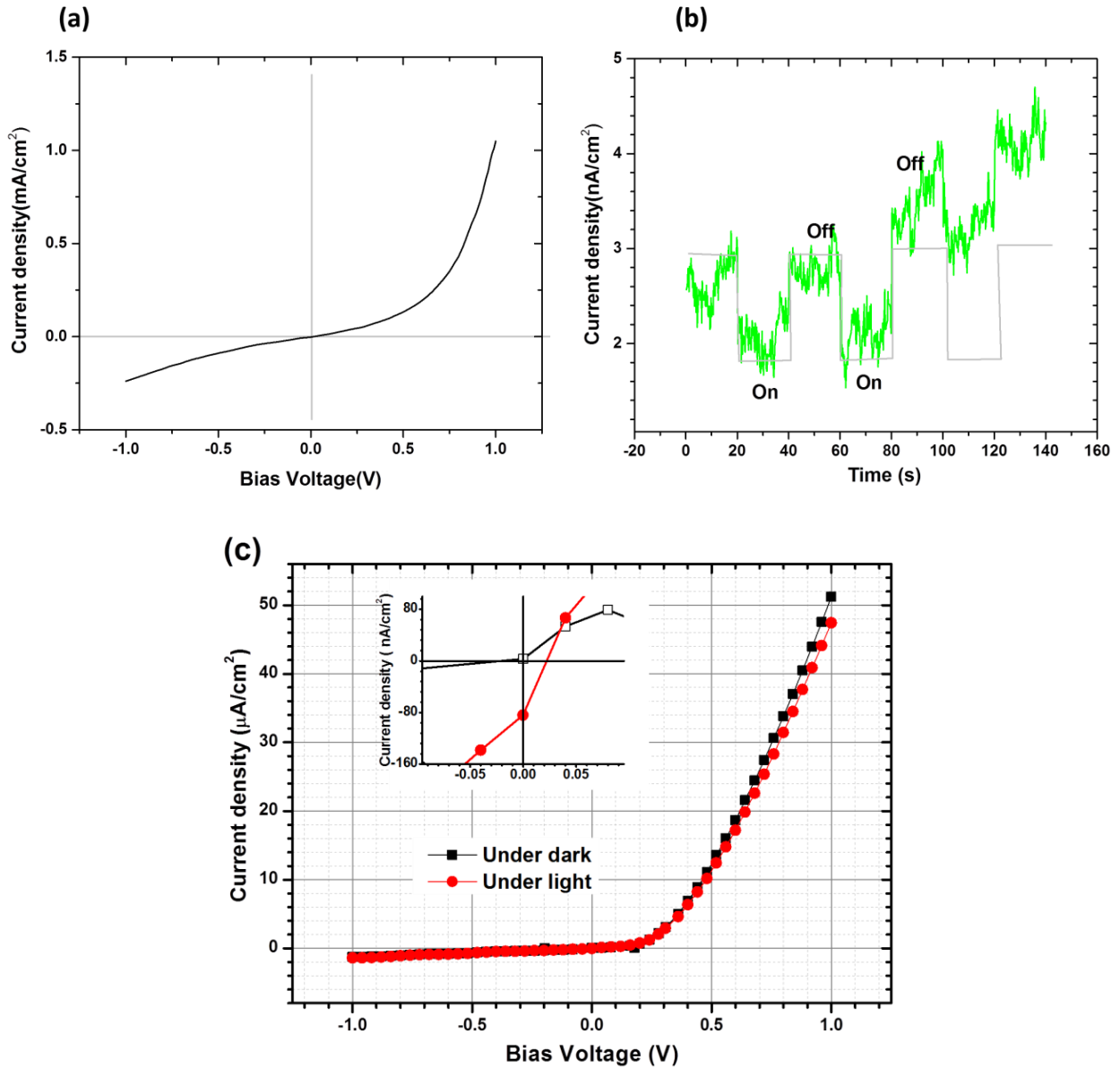

**Fig. S12**. The characteristic J-V curve for n-ZnO/p-Cu$_2$O (a, b) and n-Cu$_2$O/p-Si(111) (c) solar cells. Transient photocurrent of n-ZnO/p-Cu$_2$O cell measured under periodic LED illumination where 'On' and 'Off' step of the transient photocurrent (b) is demonstrated by the faint curve to assist the reader. A zoomed J-V curve under dark and light is shown in the inset of (c).

In Fig. S12a, the dark J-V characteristics curve of FTO/n-ZnO/p-Cu$_2$O/Au (cell#1) cell exhibited stable rectifying behaviour suggesting that a p-n junction was formed in the Cu$_2$O-ZnO system.

If p-n junction did not form, the dark and illuminated J-V curve would exhibit an ohmic behaviour and change the resistance of the system. The presence of high reverse saturation current of this cell is chiefly due to poor quality p-n junction (see Fig. S13 below). The downward shift of illuminated J-V curve was not consistent all time, therefore, taking an illuminated J-V curve within the narrow scanning range (e.g., -0.002 to +0.002V) would not produce reliable short circuit current ($J_{sc}$) and open circuit voltage ($V_{oc}$). However, the dynamic photo response at zero bias voltage of the same cell switching with the LED (wavelength just above the band gap of the absorber $Cu_2O$ ($E_g \approx 2.2eV$ (564nm)) layer) by a pulse width of 20 sec exhibited very low and noisy photocurrent yet distinguishable from the dark current (see Fig. S12b). The synchronized 'on' and 'off' photo response of the cell#1 with the LED switching confirms that $J_{sc}$ (~1 nA/cm$^2$) is real despite the noise fluctuation (see faint and green curve in figure S12b). Therefore, very low $J_{sc}$ of the PV cell#1 could be attributed to low numbers of photogenerated charge carriers owing to 'thin and poor quality' $Cu_2O$ layer (~53 nm) and thin ZnO(~114nm) layer including an amorphous interfacial layer evident from the subsequent focused ion beam (FIB) cross-section and TEM analyses (see Fig. S13 below). Referring to Fig. S12b, the dark current density is increasing over time but the amplitude of the $J_{sc}$ is remaining roughly the same on average ($J_{sc}$ ~1 nA/cm$^2$). The increase of dark current over time and low photocurrent might be attributed to the poor quality $Cu_2O$/ZnO junction owing to the presence of interfacial states which can serve as a recombination centre for the photogenerated charge carriers.

Fig. 8c shows the J-V characteristics curve of p-Si(111)/n-$Cu_2O$ (cell#2) under dark and light (The I-V curve for ohmic contact for this cell can be found in Fig. S14). A zoomed part of this J-V curve is included in the inset of Fig. 8c to distinguish low $J_{sc}$ ~ 80 nA/cm$^2$ and $V_{oc}$ ~3 mV. Though a feeble PV performance, the $J_{sc}$ and $V_{oc}$ of cell#2 composed of PLD grown n-type $Cu_2O$ on commercial p-Si substrate clearly better than cell#1. The interfacial investigation of cell#2 by FIB-TEM techniques was not done in this study which would give valuable information for further PV performance improvement.

**Interface of n-ZnO/p-Cu$_2$O (cell#1):**

To investigate the interface between ZnO and Cu$_2$O layer, a cross-section of the n-ZnO/p-Cu$_2$O(cell#1) cell was made by focused ion beam (FIB) technique. The SEM micrograph of the FIB specimen confirmed the formation of very thin but continuous Cu$_2$O (~53 nm) and ZnO (~114 nm) layer across the specimen (see figure S12 (top panel) overleaf). However, the subsequent TEM bright field (BF) image of the FIB specimen revealed an amorphous layer between the Cu$_2$O and ZnO layer (see figure S12 (bottom panel #4 inset) overleaf). It was expected that 30 min deposition time would give ~300 nm thick Cu$_2$O films, but both SEM and TEM BF image revealed a thickness which is much lower than the expected thickness. One possible reason might be the ZnO coating of quartz window occurred during ZnO film deposition step. Therefore, most of the UV-radiation might be absorbed or/and allow less amount (and less energetic) of radiation to fall on the Cu$_2$O target. This should affect the quantity of ablated species from the target as well as subsequent Cu$_2$O formation on the substrate. This means that one could expect a thin and poor quality Cu$_2$O film on any substrate, which resulted in poor quality interface between the deposited film and the substrate. A good quality Cu$_2$O film having absorption cross-section ~$10^5$ cm$^{-1}$ near the band edge suggest that at least ~100 nm thick layer is necessary for sufficient numbers of photons to be absorbed. Thicker layer is also desired to avoid short-circuiting between inter layers. Therefore, very low J$_{sc}$ of the present PV cell could be attributed to low numbers of photogenerated charge carriers owing to 'thin and poor quality' Cu$_2$O layer.

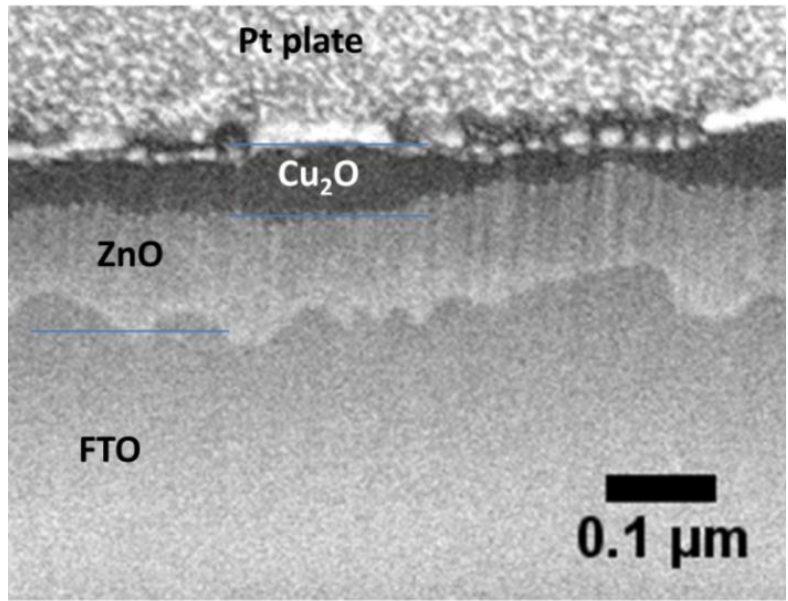
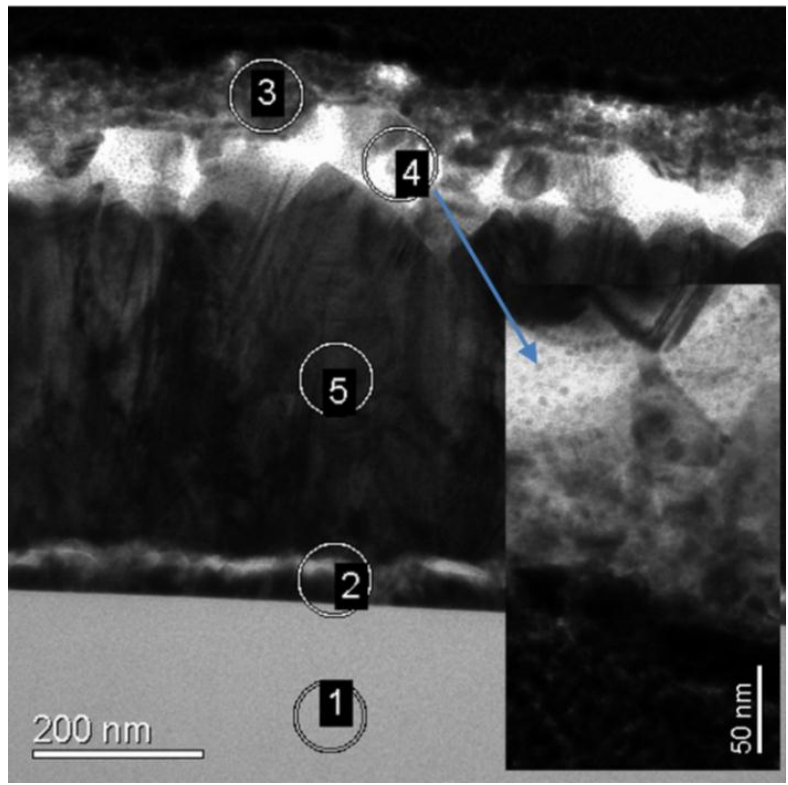

**Fig. S13.** FIB cross section of ZnO/Cu$_2$O solar cell junction (top): Cu$_2$O(~53nm), and ZnO(~114nm) layer. TEM Bright field image of the FIB cross-sectional sample (bottom left). Bottom left: Cu$_2$O(#3); amorphous layer(#4) between ZnO and Cu$_2$O layer(Shown bottom inset); FTO(#5) SiO$_2$(#2) and Glass(#1). Platinum (Pt)-plate was used to protect thin layers during FIB cross-sectional specimen preparation.

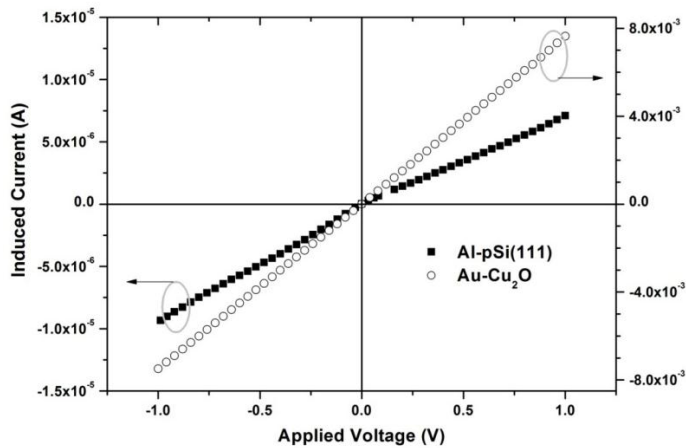

**Fig. S14.** I-V curve of p-Si (solid square) and n-$Cu_2O$ (open circle) with aluminium and gold contacts showing ohmic behaviour.

## References

[1] S.F.U. Farhad, R.F. Webster, D. Cherns, Electron microscopy and diffraction studies of pulsed laser deposited cuprous oxide thin films grown at low substrate temperatures, Materialia, 3 (2018) 230 - 238. https://doi.org/10.1016/j.mtla.2018.08.032

[2] S.F.U. Farhad, Copper Oxide Thin Films grown by Pulsed Laser Deposition for Photovoltaic Applications, in: School of Physics, University of Bristol, UK, 2016, pp. 222. https://ethos.bl.uk/OrderDetails.do?uin=uk.bl.ethos.691178

[3] Van der Pauw Hall Measurement worksheet, Physical Measurement Laboratory, National Institute of Standards and Technology (NIST). Accessed on 27/05/2020. https://www.nist.gov/pml/nanoscale-device-characterization-division/popular-links/hall-effect/sample-hall-worksheet

[4] C.M. McShane, K.S. Choi, Junction studies on electrochemically fabricated p-n $Cu_2O$ homojunction solar cells for efficiency enhancement, Physical chemistry chemical physics : PCCP, 14 (2012) 6112-6118. https://doi.org/10.1039/C2CP40502D

[5] B. Bera, A. Chakraborty, T. Kar, P. Leuaa, M. Neergat, Density of States, Carrier Concentration, and Flat Band Potential Derived from Electrochemical Impedance Measurements of N-Doped Carbon and Their Influence on Electrocatalysis of Oxygen Reduction Reaction, The Journal of Physical Chemistry C, 121 (2017) 20850-20856. https://doi.org/10.1021/acs.jpcc.7b06735

[6] K. Uosaki and and Hideaki Kita, Effects of the Helmholtz Layer Capacitance on the Potential Distribution at semiconductor/Electrolyte Interface and the Linearity of the Mott-Schottky Plot, Journal of The Electrochemical Society, 130 (1983) 895. https://doi.org/10.1149/1.2119853

[7] A. Paracchino, J.C. Brauer, J.-E. Moser, E. Thimsen, M. Graetzel, Synthesis and Characterization of High-Photoactivity Electrodeposited $Cu_2O$ Solar Absorber by Photoelectrochemistry and Ultrafast Spectroscopy, The Journal of Physical Chemistry C, 116 (2012) 7341-7350. https://doi.org/10.1021/jp301176y